\documentclass[a4paper,fleqn,usenatbib,useAMS]{mnras}

\usepackage{graphicx}	
\usepackage{amsmath}	
\usepackage{amssymb}	
\usepackage{multicol}        
\usepackage{bm}		
\usepackage{pdflscape}	

\usepackage[T1]{fontenc}
\usepackage{ae,aecompl}

\usepackage{newtxtext,newtxmath}

\title[sSFR profiles and slow quenching in the green valley] 
{SDSS IV MaNGA - sSFR profiles and the slow quenching of discs in green valley galaxies}

\author[F. Belfiore et al.] 
{Francesco Belfiore$^{1,2,3}\thanks{Email: fbelfior@ucsc.edu}$,
	Roberto Maiolino$^{2,3}$, 
	Kevin Bundy$^{1}$,
	Karen Masters$^{4}$, \newauthor
	Matthew Bershady$^{5}$, 
	Grecco Oyarz\'un$^{1}$,
	Lihwai Lin$^{6}$,
	Mariana Cano-Diaz$^{7}$, \newauthor
	David Wake$^{8, 9}$,
	Ashley Spindler$^{9}$,
	Daniel Thomas$^{4}$,
	Joel R. Brownstein$^{10}$, \newauthor
	Niv Drory$^{11}$ and
	Renbin Yan$^{12}$.
	\\
	\\$^1$ University of California Observatories - Lick Observatory, University of California Santa Cruz, 1156 High St., Santa Cruz, CA 95064, USA.
	\\$^2$ University of Cambridge, Cavendish Astrophysics, Cambridge, CB3 0HE, UK.
	\\$^3$ University of Cambridge, Kavli Institute for Cosmology, Cambridge, CB3 0HE, UK.
	\\$^{4}$ Institute of Cosmology and Gravitation, University of Portsmouth, Dennis Sciama Building, Portsmouth, PO1 3FX, UK.
	\\$^5$ University of Wisconsin - Madison, Department of Astronomy, 475 N. Charter Street, Madison, WI 53706-1582, USA.
	\\$^6$ Institute of Astronomy and Astrophysics, Academia Sinica, Taipei 106, Taiwan.
	\\$^7$ Instituto de Astronom\'\i a, Universidad Nacional Auton\'oma de M\'exico, A.P. 70-264, 04510 M\'exico, D.F., Mexico.
	\\$^{8}$ Department of Physics, University of North Carolina Asheville, One University Heights, Asheville, NC 28804, USA.
	\\$^{9}$ The Open University, Walton Hall, Milton Keynes, MK7 6AA, UK.
	\\$^{10}$ Department of Physics and Astronomy, University of Utah, 115 S. 1400 E., Salt Lake City, UT 84112, USA.
	\\$^{11}$ McDonald Observatory, The University of Texas at Austin, 2515 Speedway Stop C1402, Austin, TX 78712, USA.
	\\$^{12}$ Department of Physics and Astronomy, University of Kentucky, 505 Rose Street, Lexington, KY 40506-0057, USA.}

\pubyear{2017}

\begin{document}
	
	\pagerange{\pageref{firstpage}--\pageref{lastpage}} \pubyear{2016}
	
	\maketitle
	\label{firstpage}
	
	\begin{abstract}
	We study radial profiles in H$\alpha$ equivalent width and specific star formation rate (sSFR) derived from spatially-resolved SDSS-IV MaNGA spectroscopy to gain insight on the physical mechanisms that suppress star formation and determine a galaxy's location in the SFR-$\rm M_\star$ diagram. Even within the star-forming `main sequence', the measured sSFR decreases with stellar mass, both in an integrated and spatially-resolved sense. Flat sSFR radial profiles are observed for $\rm log(M_\star/ M_\odot) < 10.5$, while star-forming galaxies of higher mass show a significant decrease in sSFR in the central regions, a likely consequence of both larger bulges and an inside-out growth history.  Our primary focus is the green valley, constituted by galaxies lying below the star formation main sequence, but not fully passive. In the green valley we find sSFR profiles that are suppressed with respect to star-forming galaxies of the same mass at {\em all galactocentric distances} out to 2 effective radii.  The responsible quenching mechanism therefore appears to affect the entire galaxy, not simply an expanding central region. The majority of green valley galaxies of $\rm log(M_\star/ M_\odot) > 10.0$ are classified spectroscopically as central low-ionisation emission-line regions (cLIERs). Despite displaying a higher central stellar mass concentration, the sSFR suppression observed in cLIER galaxies is not simply due to the larger mass of the bulge. Drawing a comparison sample of star forming galaxies with the same $\rm M_\star$ and $\rm \Sigma_{1~kpc}$ (the mass surface density within 1 kpc), we show that a high $\rm \Sigma_{1~kpc}$  is not a sufficient condition for determining central quiescence.
	\end{abstract}
	
	
\section{Introduction} 
\label{intro}

Galaxies are observed to follow bimodal distributions in many of their fundamental properties, including star formation rate (SFR), colour, morphology and the mean age of their stellar population. Galaxy bimodality was already evident in the seminal work of \cite{Hubble1936}, and later confirmed with exquisite statistics by large-scale galaxy surveys like the Sloan digital sky survey (SDSS, \citealt{Strateva2001, Kauffmann2003a, Blanton2009}). These studies motivated the division of galaxies in the colour-magnitude diagram into the star forming `blue cloud' and the passive `red sequence'. 

Thanks to its superior sensitivity to young hot stars, UV photometry from the \textit{GALEX} satellite has convincingly demonstrated the existence of a third population of galaxies lying at intermediate colours, in the so-called `green valley' (GV, \citealt{Wyder2007, Martin2007}). Since its discovery, the GV has been widely interpreted as a cross-road in galaxy evolution, being populated by galaxies in transition. This interpretation is supported by the fact that the mass flux through the GV is roughly comparable to that needed to assemble the red sequence \citep{Martin2007}, and that GV objects are intermediate in terms of other physical properties, such as S\'ersic index/concentration and stellar population ages \citep{Schiminovich2007, Mendez2011, Pan2013, Pandya2017a}. As expected from their intermediate colours, GV galaxies lie mostly below the star formation main sequence (SFMS), the tight relation in the $\rm M_\star$ - SFR plane inhabited by star forming galaxies \citep{Noeske2007, Salim2007, Renzini2015}. 

While GV galaxies host more massive bulges and are more concentrated than their blue cloud counterparts \citep{Schiminovich2007, Fang2013}, it remains unclear to which extent galactic subcomponents such as the disc and the bulge contribute to the suppression of star formation (`quenching'\footnote{We acknowledge that there is no agreement in the community with regards to the correct use of the term `quenching'. In this work we refer to quenching as the process that generates the population of galaxies below the SFMS in the $\rm M_\star$ - SFR plane. This definition has the advantage of relying entirely on observables, easily accessible at a variety of redshifts. Alternatively quenching may also be defined as a sharp break in the star formation history of a galaxy, generally implying a rapid transition to quiescence. We find this second definition more problematic, especially in light of the difficulty in deriving star formation histories and the possibility of a slow transition to quiescence.}) in this population. \cite{Dressler2014} recently suggested that the intermediate colours and depressed sSFR typical of the GV are a natural consequence of the coexistence of red, passive bulges and blue, star forming discs. According to this scenario the green valley may be explained  by a change in bulge to total mass ratio, since the contribution of the (red) bulge increases relative to the (blue) disc moving away from the SFMS towards the GV. This interpretation of the GV as a bulge-disc `purple' composite is echoed in the work of \cite{Abramson2014}, who suggest that galaxy discs form stars at constant sSFR, when the mass of the passive bulge is correctly accounted for. However, other authors disagree with this claim, finding that even when bulges are taken into account galaxy discs do not show a constant sSFR as a function of total mass \citep{Guo2015a, Whitaker2015, Schreiber2016}.

More detailed analysis of the optical colours of bulge and discs in SDSS is presented by \cite{Morselli2017}. They find that mean disc $g-r$ colours get redder when moving from the SFMS towards the quiescent region, pointing to a systematic difference in the star formation rate properties between SFMS and GV discs. They also find a reddening of the disc in higher-mass systems, which may suggest a decrease in the sSFR of discs on the SFMS.
	
These works are based on either broad-band photometry or spatially-limited spectroscopy, and may therefore suffer from systematics related to dust extinction and uncertain aperture corrections. Recently \cite{Catalan-Torrecilla2017} presented a first study of the star formation properties of discs, bars and bulges using integral field spectroscopy (IFS) from the CALIFA survey \citep{Sanchez2012}. They traced star formation spectroscopically and confirmed that disc sSFR decreases with stellar mass. Their work, however, did not specifically address the GV galaxy population.

In \cite{Belfiore2017} we have demonstrated, using IFS from the MaNGA survey \citep{Bundy2015}, that a large fraction of galaxies in the GV ($\sim 40 \%$) have quiescent central regions, while hosting star formation in extended outer discs. The central regions in these sources are not devoid of line emission, but are characterised by low H$\alpha$ equivalent widths and line ratios typical of low ionisation emission-line regions (LIERs). The line emission is generally not due to an active galactic nucleus (AGN), but is caused by UV radiation emitted by hot evolved stars \citep{Trinchieri1991, Binette1994, CidFernandes2011, Belfiore2016a}. 

In this work we exploit IFS data from the MaNGA survey to study the equivalent width of H$\alpha$ [EW(H$\alpha$)] and sSFR profiles of nearby galaxies on the SFMS and the GV and assess the relative importance of quiescent central regions and star forming discs. In particular, we study whether the intermediate sSFR of GV galaxies is due their being composite `purple' systems, with quiescent bulges and star forming discs, or whether discs form stars with different sSFRs depending on their galaxy host. 
This work is part of a series of papers from the MaNGA team dedicated to mapping sSFR on resolved scales. \cite{Spindler2018} focus on the effect of environment on the sSFR profiles, while \cite{Sanchez2017} study sSFR profiles in AGN hosts. In \cite{Lin2017c} we present the sSFR properties for a sample of three green valley galaxies observed by MaNGA and ALMA with matched resolutions. Finally, \cite{Ellison2017} present an independent analysis of sSFR profiles of MaNGA galaxies, with emphasis on the spatially-resolved relation between SFR and $\rm M_\star$ surface densities.

We start by describing the MaNGA IFS data used in this work and the sample selection in Sec. \ref{data}. In Sec. \ref{results} we analyse the EW(H$\alpha$) and sSFR profiles for SFMS and GV galaxies, and the role of quiescent central regions. In Sec. \ref{dis} we discuss these results and their influence on our understanding of the quenching of star formation in the GV.

\section{Data and sample} 
\label{data}

\subsection{The MaNGA data}
\label{sec2.1}

The MaNGA survey \citep{Bundy2015, Yan2016a}, part of SDSS-IV \citep{Blanton2017}, aims to obtain spatially resolved spectroscopy for a representative sample of 10~000 galaxies in the redshift range 0.01 $<$ z $<$ 0.15 by 2020. The MaNGA instrument operates on the SDSS 2.5m telescope at Apache Point Observatory \citep{Gunn2006} and consists of a set of 17 hexagonal fibre bundles of different sizes, plus a set of mini-bundles and sky fibres used for flux calibration and sky subtraction respectively \citep{Drory2015, Law2015, Yan2016}. All fibres are fed into the dual beam BOSS spectrographs covering the wavelength range from 3600 \AA\ to 10300 \AA\ with a spectral resolution R $\sim$ 2000 \citep{Smee2013}. 

MaNGA galaxies are selected from an extended version of the Nasa-Sloan (NSA) catalogue and are observed out to 1.5 $\rm R_e$ (primary sample, comprising 2/3 of the total sample) or 2.5 $\rm R_e$ (secondary sample, comprising 1/3 of the total sample). Both subsamples are independently selected to be representative of the overall galaxy population at each stellar mass in the range $\rm 9.0 < log(M_\star/M_\odot) < 11.0$ (in practice $M_i$ is used for sample selection to avoid the systematic uncertainty intrinsic in deriving stellar masses, \citealt{Wake2017}).

The MaNGA data used in this work was reduced using version {\tt v1\_5\_1} of the MaNGA reduction pipeline \citep{Law2016a}. 
Our starting sample consists of all MaNGA galaxies observed within the first $\sim$ 2 years of operation, corresponding to the publicly available SDSS data release 13 (DR13, \citealt{SDSS_DR13}), which includes 1352 unique galaxies. 

\begin{figure} 
	\centering
	\includegraphics[width=0.45\textwidth, trim=20 0 30 30, clip]{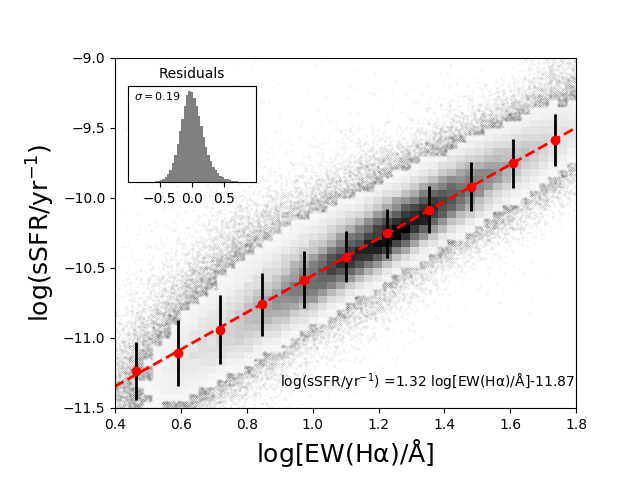}
	\caption{A 2D histogram of the relation between sSFR and EW(H$\alpha$) in emission for all the spaxels classified as star forming using the [SII]/H$\alpha$ BPT diagram in  the sample of MaNGA galaxies considered in this work. For bins with less than 100 spaxels the individual data points are shown. Red points with black error bars show the median and median absolute deviation in bins of log[EW(H$\alpha$)]. The red dotted line is a linear fit to the median relation. The derived parameters of the linear fit are also shown. The inset histogram shows the residuals from the best-fit relation, indicating a scatter of 0.19 dex.}
	\label{fig1}
\end{figure}

\subsection{Spectral fitting}
\label{sec2.2}

Physical parameters of the continuum and the emission lines are obtained via a customised spectral fitting procedure described in \citealt{Belfiore2016a}, with some differences detailed below. 
For each galaxy, the stellar continuum is binned to a minimum signal to noise ($\rm S/N=6$)  and fitted using a set of MIUSCAT simple stellar population models \citep{Vazdekis2012} including 9 ages (from 0.063 to 17.8 Gyr) and 4 metallicities ($\rm [Z/H]=-0.71$ to 0.22).
The fit is performed with penalized pixel fitting \citep{Cappellari2004}. Regularisation is employed in order to derive the smoothest star formation history which is consistent with the data, as described in \cite{Cappellari2017}.  We fit the stellar continuum with a \cite{Calzetti2000} extinction law and no additive polynomials. Only 0.1\% of spaxels which have continuum S/N$>$6 require $\rm E(B-V) >$0.5, despite the code allowing $\rm E(B-V)$ values as large as 10. We note that the continuum reddening is not tied to the Balmer decrement, which is computed independently in a subsequent step.
The stellar mass for each resolved region within the galaxy is calculated as the sum of the weights for each simple stellar population, assuming a \cite{Chabrier2003} initial mass function.  

After subtracting the stellar continuum the emission lines are fitted on a different binning scheme. To increase the ability to fit weaker lines the velocities of all lines are tied together. In this way we effectively use the stronger lines to constrain the kinematics of the weaker ones. 
The reddening of the emission lines is calculated from the Balmer decrement, using the $\rm H\alpha/H\beta$ ratio and a \cite{Calzetti2000} attenuation curve with $\rm R_V = 4.05$.
The theoretical value for the Balmer line ratio is taken from \cite{Osterbrock2006}, assuming case B recombination ($ \rm H\alpha/H\beta=2.87$). In order to obtain a reliable extinction correction we select only spaxels with S/N $>$ 3 on both H$\alpha$ and H$\beta$. We have verified that the E(B-V) values for line emission and continuum correlate with each other and that, as expected, the E(B-V) for the gas is larger than that of the continuum.

\subsection{Measuring EW(H$\alpha$) and sSFR}
\label{sec2.2e}
The EW of the nebular lines is measured by dividing the line flux (with no extinction correction) by a measure of the local continuum. For simplicity of notation, we define equivalent widths to be positive in emission. The limiting detectable EW depends on the relative ratio of the S/N on the line and on the continuum, and on the velocity dispersion of the line \citep{Sarzi2006}. For a barely detected H$\alpha$ line (S/N $=$ 2) with typical dispersion ($\rm \sigma_{H\alpha}= 80 \ km \ s^{-1}$, close to the instrumental resolution), the limiting EW is $\sim 1.5 \AA$ for a continuum S/N $=$ 6. Lower EWs are routinely detected in MaNGA in spaxels where the continuum S/N is higher.

The SFR is calculated from extinction-corrected H$\alpha$ using the conversion formula from \cite{Kennicutt1998} and a \cite{Chabrier2003} initial mass function. We note that the use of the \cite{Salpeter1955} initial mass function would imply SFR higher by a factor of 0.2 dex, but also an increase in stellar mass by a similar factor (0.22 dex, \citealt{Madau2014}), eventually resulting in nearly identical sSFR values.

We classify each spaxel within a galaxy according to the \cite{Kewley2001} demarcation line in the [SII]$\lambda\lambda$6717,31/H$\alpha$ ([SII]/H$\alpha$) versus [OIII]$\lambda$5007/H$\beta$ Baldwin-Phillips-Terlevich (BPT) diagnostic diagram \citep{Baldwin1981, Veilleux1987, Kauffmann2003a, Kewley2006, Belfiore2016a} as either star forming, LIER of Seyfert-like. For spaxels classified as star forming in the BPT diagram we assume that all the H$\alpha$ flux is directly associated to star formation. For LIER-like spaxels we assume that the H$\alpha$ flux is a sum of a component due to star formation and a `pure' LIER component unrelated to star formation. Following \cite{Blanc2009} and \cite{Belfiore2017} we make use of the [SII]/H$\alpha$ ratio to quantify the residual star formation in LIER-like regions. In detail, if we denote the fraction of H$\alpha$ flux contributed by star formation and pure LIER-like emission as $\rm f_{SF}$ and $\rm f_{L}$ respectively, we have 
\begin{equation}
\rm f_{SF}+ f_{L}=1, \\ 
\left( \frac{[SII]}{H\alpha} \right)= f_{SF} \left( \frac{[SII]}{H\alpha} \right)_{SF}+ f_{L} \left( \frac{[SII]}{H\alpha} \right)_{L}
\label{eq_correct}
\end{equation}
We assume typical line ratios of $\rm \left( \frac{[SII]}{H\alpha} \right)_{SF}=0.4$ and $\rm \left( \frac{[SII]}{H\alpha} \right)_{L}=1.0$ from the analysis of  \cite{Belfiore2016a}, but we have checked that a change in 0.1 dex in these line ratios does not significantly affect the conclusions of the paper. We note that, in reality, the [SII]/H$\alpha$ ratio will depend on both metallicity and ionisation parameter, and therefore the correction presented here may be inaccurate for galaxies on an individual basis. Ideally one would study the whole mixing sequence (e.g. \citealt{Davies2014}) to determine ([SII]/H$\rm \alpha)_{SF}$ and  ([SII]/H$\rm \alpha)_{L}$ separately for each galaxy, but this is unfeasible due to the limited spatial resolution of the MaNGA data. Finally, we note that this procedure, which assigns a non-zero SFR to LIER spaxels, does not substantially affect the integrated SFR values. It has some impact, however, on the radial profiles of central LIER galaxies, as discussed in Sec. \ref{sec3.2}.

Using repeat observations, \cite{Yan2016a} demonstrate that for $\rm log(SFR/M_\odot \ yr^{-1} \ kpc^{-2}) > -2.7$ (converting their values to a Chabrier IMF) and $\rm E(B-V)< 0.5$, the uncertainty in the SFR is less than 0.2 dex. This uncertainty is dominated by the uncertainty in the line fluxes and not by the absolute and relative spectrophotometric calibration \citep{Yan2016}. Using the current MaNGA data, we obtain that a S/N $=$ 3 for H$\beta$ corresponds to a median $\rm log(SFR/M_\odot \ yr^{-1} \ kpc^{-2}) \sim - 3.0$. We take this to represent the median SFR sensitivity limit in MaNGA, but in the case of low extinction and for the lowest redshift objects, lower SFR can be measured. 

In Fig. \ref{fig1} we show the relationship between EW(H$\alpha$) and sSFR for all the star forming spaxels in the current MaNGA sample. 
The relationship between sSFR and EW(H$\alpha$) is well-fitted by a power law over almost 2 dex in sSFR, given by 

\begin{equation}
\rm log(sSFR/yr^{-1})=-11.87 + 1.32 \ log(EW(H\alpha) / \text{\normalfont\AA}). 
\end{equation}
The formal errors on the best fit coefficients are less than 1\% due to the large number of datapoints. Our best fit coefficients agree well with those reported by \cite{Sanchez2013} in a study based on CALIFA data. The relation between sSFR and EW(H$\alpha$) is superlinear, possibly due to the M/L dependence of the EW. In fact, EW/sSFR $\rm \sim (F_{H\alpha}/ L_r) (M_\star / SFR)  \sim (M_\star/L_r)$ where $\rm L_r$ is the light emitted in the continuum passband around H$\alpha$. Higher EW(H$\alpha$) corresponds to younger stellar populations and hence lower M/L, going in the direction of the observed trend. Differences in relative extinction between gas and the stellar component will also affect the relation between sSFR and EW(H$\alpha$) \citep{Calzetti1994, Calzetti2000}. 

This work focuses on the discussion of sSFR profile, but we use the EW(H$\alpha$) as a cross-check to demonstrate that our conclusions are robust to the modeling assumptions (e.g. the derivations of the M/L for the stellar population and the dust extinction correction) employed in deriving the sSFR.

\subsection{The galaxy sample}
\label{sec2.3}

Following the classification scheme proposed in \cite{Belfiore2016a}, we subdivide galaxies according to their excitation morphologies in star forming (SF, dominated by star formation at all radii), central LIER (cLIER, LIER emission in the central regions but star forming at larger galactocentric distances), extended LIER (eLIER, LIER emission at all radii), line-less (no line emission detected) and Seyfert (Seyfert line ratios in the central 3$''$).  Operationally, galaxies are classified as cLIERs if their line emission from a central aperture of 3$''$ in diameter is classified as LI(N)ER in the [SII]-BPT diagram using the demarcation lines of \cite{Kewley2006}, but star forming regions are also present at larger galactocentric distance. As already noted in \cite{Belfiore2016a}, all galaxies in MaNGA hosting star formation have detected line emission in the bulge/central regions, showing either star forming, Seyfert or LIER line ratios. This means that there exist no line-less bulges in the local Universe, unless the whole galaxy is line-less.

In this work we are interested in both star forming and transitioning GV galaxies with residual star formation. We therefore limit our study to the spectroscopically classified star forming and cLIER galaxies (814). We note that Seyfert galaxies may host star formation but they are excluded in this work, as they are the subject of a detailed parallel study \citep{Sanchez2017}. 

For eLIER galaxies an integrated SFR could be computed making use of equation \ref{eq_correct}, but we refrain from doing so because it is fundamentally difficult to measure low levels of star formation in these systems. Our approach for calculating the SFR of LIERs assumes that [SII]/H$\alpha$ line ratios intermediate between those typical of star formation and LIER emission are due to a mixing sequence between the two ionisation sources. This scenario may not apply to eLIER galaxies, which have generally flat [SII]/H$\alpha$ gradients \citep{Belfiore2016a} and where local changes in [SII]/H$\alpha$ may be due to physical processes unrelated to star formation (e.g. shocks). If we nonetheless computed a SFR for eLIERs using equation \ref{eq_correct}, the resulting SFR would lie more than 2 dex below SFMS derived in the next subsection (see for example Fig. 4 of \citealt{Belfiore2017}), confirming that these galaxies are fully quenched.

For a reliable study of radial gradients we impose further cuts on our galaxy sample by requiring the major to minor axis ratio (b/a) to be greater than 0.4 (to exclude high-inclination systems, leaving 701 galaxies), $\rm 9.0 < \log(M_\star/M_\odot) < 11.5$ (651) and excluding visually-classified mergers or closely interacting systems, leading to a final sample of 559 galaxies.

\begin{figure*} 
	\centering
	\includegraphics[width=0.65\textwidth, trim=0 0 30 30, clip]{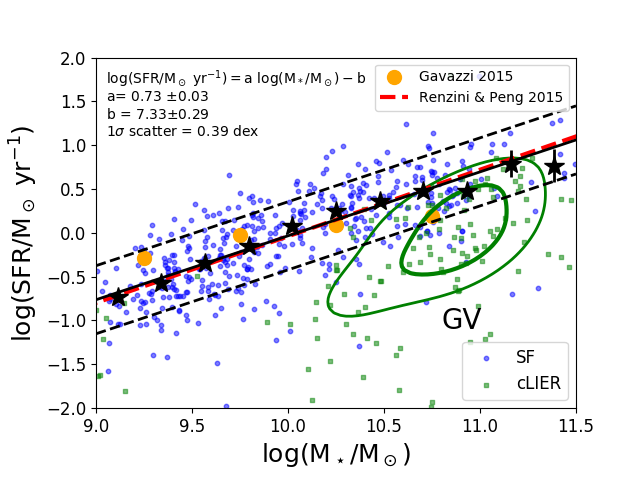}
	\caption{The position of star forming (blue circles) and cLIER (green squares) galaxies in the SFR versus $\rm M_\star$ diagram. The SFR and $\rm M_\star$ values are obtained from the MaNGA data integrating within a $\rm 2.5 \ R_e$ aperture. An aperture correction is performed for MaNGA galaxies not observed out to $\rm 2.5 \ R_e$. 
	The black stars represent the bi-weight (a robust estimate for the mean) SFR values in 11 equally spaced mass bins for the star forming sample. The black solid line represents a linear fit to these values, and the scatter around the fitted relation is shown by the black dashed lines. In this work galaxies lying more than 1$\sigma$ below the SFMS are considered to belong to the green valley (GV), while main sequence galaxies are taken to lie within the scatter and above the relation. 
	Determinations of the SFMS from the literature are also plotted.  Solid orange circles represent the mean SFR values in $\rm M_\star$ bins from \protect\cite{Gavazzi2015}, while the red dashed line (partially overlapping the black solid line) represents the best fitting SFMS from \protect\cite{Renzini2015}.} 
	\label{fig2}
\end{figure*}

\subsection{The integrated star formation main sequence}
\label{sec2.3.1}

The GV in the nearby Universe is traditionally defined as a locus in colour-magnitude space, employing $NUV -r $ or $g-r$ colours.
While $NUV- r$ selection has the advantage of being model-independent at $\rm z=0$, the use of a single colour fails at classifying galaxies at higher redshift due to the larger fraction of dusty objects \citep{Williams2009, Wild2014a}. Moreover, even at $\rm z=0$ it remains difficult to correct the $NUV$ flux for the effects of dust extinction in absence of detailed multi-band photometry, which is not available for the MaNGA targets. In this paper we therefore take a more physical approach and define the GV with respect to the SFMS. While star formation rates and stellar masses are affected by modeling systematics, the main advantage of this choice lies in the fact that the relative position of galaxies with respect to the SFMS can generally be measured robustly, both in observational datasets and in galaxy evolution models. 

In this work we define the GV as the locus in the SFR-$\rm M_\star$ plane which lies more than 1$\sigma$ below the SFMS (where $\sigma$ is the measured scatter across the SFMS). We refer to galaxies lying within the scatter of the SFMS or above the relation as `main sequence galaxies'. In Appendix \ref{app1} we discuss how this definition of the GV compares to more traditional colour-based determinations. In summary, we find that our definition is closely equivalent to the selection obtained using an $NUV - r$ colour cut, while $g-r$ colours perform much worse at isolating galaxies below the SFMS.

The determination of the (integrated) SFMS can be potentially affected by several systematic effects. Here we discuss two of them: the selection of the `star forming' population and the correction for aperture effects. Regarding the first one, several authors have explicitly pre-selected galaxies according to their colour or sSFR before including them in the SFMS (e.g. \citealt{Noeske2007, Peng2010, Guo2015a}). The effect of this selection is most apparent at high masses, where there are few star forming blue galaxies and the exact nature of the pre-selection may lead to varying amounts of contamination by the transition/passive population. In this work we choose to define the SFMS with respect to the population of BPT-classified star forming galaxies, excluding cLIERs. We comment below on the effect of this decision on the derived SFMS, and its slope at high masses.

Secondly, determinations of the SFMS from spectroscopic data are generally affected by aperture biases, due to the finite extent of the spectroscopic coverage \citep{Brinchmann2004}. The MaNGA IFS data represents a substantial improvement over single-fibre SDSS spectra, but nonetheless presents us with an heterogeneous set of apertures, since different galaxies are covered out to different galactocentric distances. In particular, the MaNGA primary and secondary subsamples are covered to at least $\rm 1.5 \ R_e$ and $\rm 2.5 \ R_e$ respectively, although many galaxies in both subsamples are covered out to larger radii. 

In order to derive an aperture-free SFMS in this work we compute the integrated SFR and $\rm M_\star$ within a $\rm 2.5 \ R_e$ aperture. When the MaNGA IFU does not fully cover a galaxy out to $\rm 2.5 \ R_e$ we employ an aperture correction procedure based on the secondary sample. We are justified in using the secondary sample for this procedure because this subsample is independently selected to be representative of the local galaxy population and is \textit{not} constituted by a biased collection of smaller or more distant galaxies. The aperture correction procedure and a detailed comparison of the resulting stellar masses with previous determinations from the literature are presented in Appendix \ref{app2}. The aperture corrections used in this work are overall very small (an average of 6\% for both integrated stellar masses and the SFR) and not applying them has no impact on our results.

In Fig. \ref{fig2} we show the position of star forming and cLIER galaxies used in this work in the SFR-$\rm M_\star$ plane, where SFR and $\rm M_\star$ are derived within a $\rm 2.5 \ R_e$ aperture. We define the SFMS as the linear relation between $\rm log(SFR)$ and $\rm log(M_\star)$ obtained by considering only the star forming galaxies and fitting the median SFR in $\rm M_\star $ bins (values shown as black stars in Fig. \ref{fig2}). The derived relation has sub-linear slope and is given by
\begin{equation}
\rm log(SFR/M_\odot \ yr^{-1})= (0.73 \pm 0.03) \  log(M_\star/M_\odot) - (7.33\pm 0.29). 
\end{equation}
This linear relation and its scatter (0.39 dex) is shown in Fig. \ref{fig2} by solid black and dashed lines respectively.

Our results are in good agreement with the \cite{Renzini2015} SFMS determination based on SDSS data (dashed red line in Fig. \ref{fig2}, partially overlapping with the best fit SFMS obtained in this work). \cite{Gavazzi2015} studied the SFMS in the local Universe based on H$\alpha$ narrow-band imaging and extended the relation to low stellar masses $\rm log(M_\star/M_\odot)=[7.0, 9.0]$. They find a change in slope from near unity at low masses to a shallow power law (of exponent 0.2) above $\rm log(M_\star/M_\odot)=9.45$. Their median SFR values in mass bins for $\rm log(M_\star/M_\odot) > 9.0$ are shown as orange circles in Fig. \ref{fig2} and are consistent with our SFMS determination within 1$\sigma$. 

Interestingly, a flattening in the slope of the SFMS is obtained using our dataset if galaxies with quiescent central regions (cLIERs) are included in the fit. This may bring the results of this work in closer agreement with those of \cite{Gavazzi2015}, who do not remove centrally quiescent galaxies in their analysis, pointing towards the importance of a consistent pre-selection of galaxies when comparing different determinations of the SFMS.

cLIER galaxies are found at the high-mass end of the SFMS and across the GV \citep{Belfiore2017}, as indicated by the green density contour in Fig. \ref{fig2}. In Appendix \ref{app3} we show that central LIER emission is closely linked to other possible definitions of quiescence, like $\rm D_N(4000)$. Interestingly, cLIERs are virtually absent for $\rm M_\star/ M_\odot < 10^{10.0}$, except for a small population of 9 galaxies. Visual inspection of the emission line maps of these objects reveals that these galaxies have irregular morphologies and host clumpy line emission. In cases where the galaxy centre does not correspond to an area of high star formation, it may be dominated by diffuse ionised gas, thus making these galaxies appear as central LIER objects. In light of this, we exclude these galaxies from the cLIER sample. The final number of SFMS and GV, as well as star forming and cLIER galaxies used in this work is reported in Table \ref{gal_properties}.

Since MaNGA observes more massive galaxies at higher redshift, the cosmic evolution of the sSFR may have some influence on our results. The median redshift of star forming galaxies in the mass bin $\rm log(M_\star/M_\odot) = [9.0-9.5]$ is $\rm <z>=$ 0.028 while for the mass bin  $\rm log(M_\star/M_\odot) = [11.0-11.5]$ bin is $\rm <z>=$ 0.088. If we assume the sSFR($\rm M_\star, z$) relation from \cite{Leitner2012}, based mostly on the data from \cite{Karim2011}, the sSFR in our highest mass bin is overestimated by 0.08 dex compared to the lowest mass bin. There is also a redshift difference between the primary and secondary sample, but its effect of the derived sSFR is negligible.
Considering the varying prescriptions for sSFR($\rm M_\star, z$) present in the literature, we feel that the systematic uncertainty in the correction is larger than the correction itself. We therefore do not correct our sSFR values for the redshift dependence (however median redshifts for each stellar mass bin are provided in Table \ref{table2} below, should the reader wish to apply a correction). We note that, since the correction would further decrease the sSFR of high-mass galaxies, it would imply an even shallower slope for the SFMS. 

\begin{table}
	\caption{Sample of MaNGA galaxies used in this work. Note that 9 galaxies with $\rm log(M_\star/M_\odot)< 10.0$ are excluded from the cLIER sample as described in the text.}
	\label{gal_properties}
	\centering
	\begin{tabular}{ l c  }
		Type & N. galaxies \\
		\hline
		Sample used in this work & 559 \\
		\hline 
		main sequence (SFMS) &  407 (73\%) \\
		Green Valley (GV, >1$\sigma$ below SFMS) & 152 (27\%) \\
		\hline 
		star forming (SF) &  437 (79\%)\\
		central LIER (cLIER) & 113 (21\%) \\
		\hline
	\end{tabular}
\end{table}

\section{sSFR and EW(H$\alpha$) profiles}
\label{results}

We make use of elliptical Petrosian effective radii ($\rm R_e$) and inclinations from the NSA catalogue to construct de-projected radial gradients, following the procedure utilised in \cite{Belfiore2017a}. In summary, elliptical annuli with semi-major axis of 0.15 $\rm R_e$ are constructed and the median EW(H$\alpha$), $\rm \Sigma_{SFR}$ (SFR surface density) and $\rm \Sigma_\star$ (stellar mass surface density) are computed for each annulus. Regions where star formation is not detectable (including regions with low S/N or $\rm f_{SF}=0$) are considered as having zero star formation in the averaging procedure. The $\rm \Sigma_{sSFR}$ (sSFR surface density) profile is then obtained by dividing the $\rm \Sigma_{SFR}$ by the $\rm \Sigma_\star$ in each annulus.

\begin{figure*} 
	\includegraphics[width=0.95\textwidth, trim=30 0 40 0, clip]{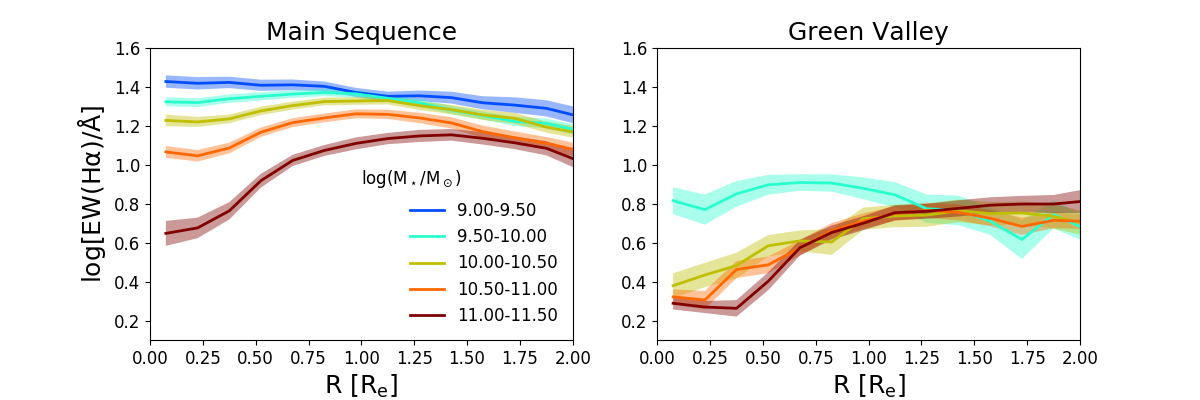}
	\includegraphics[width=0.95\textwidth, trim=30 0 40 0, clip]{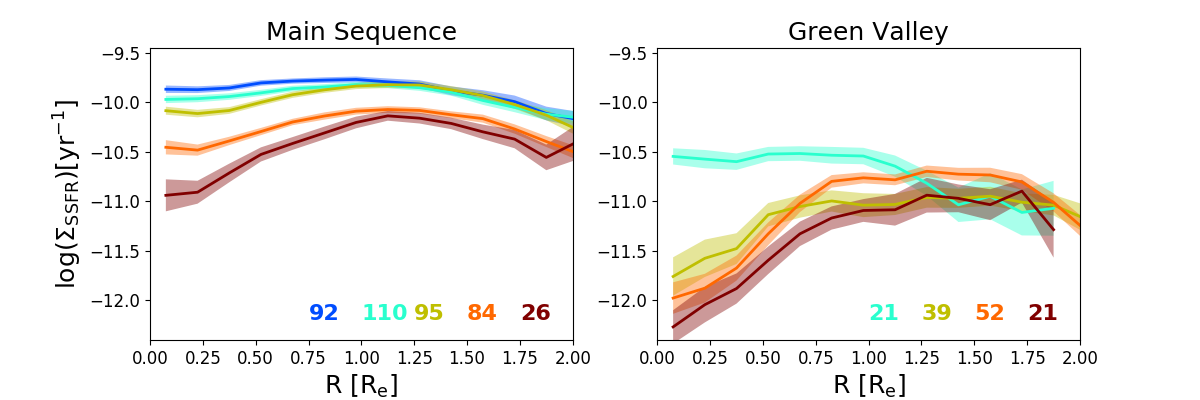}
	
	\caption{\textit{Top}: The stacked EW(H$\alpha$) radial profiles for different mass bins on the SFMS (left) and in the GV (right). Profiles are computed by deprojecting the gradient using the photometric inclinations from the NSA catalogue and normalised to the elliptical Petrosian effective radius ($\rm R_e$). Elliptical annuli of 0.15 $\rm R_e$ are used in the stacking procedure.
	\textit{Bottom}: The stacked sSFR radial profiles for different mass bins on the SFMS (left) and in the GV (right). The sSFR computed as the ratio of the median $\rm \Sigma_{SFR}$ and the $\rm \Sigma_\star$ in each annulus. Regions with undetectable star formation (either due to low S/N or whose line emission is classified as fully LIER) are assigned zero SFR in the averaging procedure. SFR of regions in the transition zone between the BPT-classified star forming and LIER sequences is computed according to equation \ref{eq_correct}. The number of galaxies in each mass bin is shown in the bottom right corner. Only mass bins containing more than 20 galaxies are shown.} 
	\label{fig3}
\end{figure*}

\subsection{The SFMS and the green valley}
\label{sec3.1}

The EW(H$\alpha$) and sSFR profiles for SFMS and GV galaxies in five stellar mass bins, going from $\rm log(M_\star/M_\odot)= 9.0$ to $\rm log(M_\star/M_\odot)=11.5$ in 0.5 dex intervals, are presented in Fig. \ref{fig3}. For each stellar mass bin, the radial profile is computed as the Tukey biweight of the profiles of individual galaxies. Error bars are obtained by calculating a robust estimator for the sample standard deviation and dividing by $\sqrt{N}$, where N is the number of profiles at each radius.

For SFMS galaxies we observe a regular change in the shape of the EW(H$\alpha$) and sSFR radial profiles with stellar mass. Both EW(H$\alpha$) and sSFR profiles show two notable features: a sharp, strongly mass-dependent decrease in the central regions, and a slow decrease in the sSFR profile at large radii, with a milder mass dependence.

In the innermost radial bin, the sSFR difference between the most massive and the least massive stack is as large as 1.1 dex. Moreover the most massive bin reaches sSFR values as low as $\rm log(sSFR/yr^{-1})=-11.0$ in the innermost regions. To put the absolute scale of the sSFR into context, a useful number to keep in mind is the characteristic sSFR for a main sequence galaxy of $\rm M_\star = 10^{10} \ M_\odot$ in SDSS, which is $\rm log(sSFR/ yr^{-1}) \sim -10$ \citep{Peng2010, Renzini2015}. 

The key result of this work, evident from Fig. \ref{fig3}, is the existence of a systematic difference in the sSFR [and EW(H$\alpha$)] profiles between the SFMS and the GV at all radii. In each stellar mass bin, EW(H$\alpha$) and sSFR are suppressed in the GV with respect to the SFMS \textit{at all radii}. At 1.0 $\rm R_e$, for example, the suppression in sSFR between the SFMS and the GV median profiles is 0.6-1.2 dex, depending on the mass bin considered. At 2.0 $\rm R_e$ the median suppression is between 0.6-1.0 dex.

The main differences in the EW(H$\alpha$) and sSFR profiles are preserved if, instead of using the SFMS, galaxies are classified according to their position in $ NUV-r $ colour diagram. We have also checked that the variable spatial coverage of the two MaNGA subsamples does not bias the results at large radii. In fact, the different radial coverage of the primary and secondary samples may lead to a bias in the stacked line profiles because for $\rm R> 1.5~R_e$ the fraction of primary sample covered may represent a biased subset of the overall galaxy population. To address this bias we have computed the profiles obtained using only the secondary sample (roughly one third of the total sample). We find that the qualitative features of the profiles at large radii are preserved with this smaller subsample, confirming that the combination of the two subsamples is not biasing the overall result.

\subsection{The role of centrally quiescent regions}
\label{sec3.2}

\begin{figure} 
	\centering
	\includegraphics[width=0.45\textwidth, trim=0 0 30 0, clip]{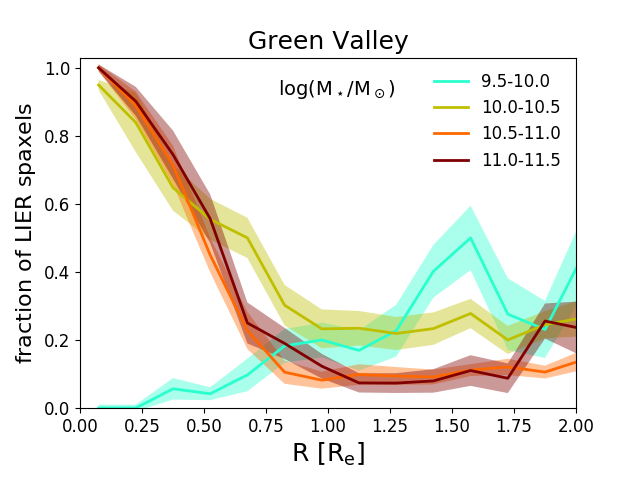}
	\caption{The fraction of spaxels classified as LIERs (according to the [SII]-BPT diagram) in GV galaxies of different stellar masses as a function of galactocentric radius, normalised to the effective radius. GV galaxies are dominated by LIER emission for $R< 0.5\  R_e$ and $\rm log(M_\star/M_\odot)> 10.0$.} 
	\label{fig4}
\end{figure}

The predominance of LIER emission is partly responsible for the low values of sSFR and EW(H$\alpha$) observed in the central regions of massive galaxies in the green valley (and, to a lesser extent, in the SFMS). To quantify the impact of LIER quiescent regions in the GV, in Fig. \ref{fig4} we show the fraction of spaxels classified as LIERs as a function of radius for GV galaxies of different stellar masses. The figure demonstrates the increasing predominance of LIER emission with decreasing galactocentric radius and mass. GV galaxies are typically dominated by LIER ionisation (i.e. more than 50\% LIER fraction) for $\rm R< 0.5 \ R_e$ and $\rm log(M_\star/M_\odot)>10.0$, while lower mass galaxies show almost no instances of LIER emission in their central regions.

The increasing importance of LIER regions is also evident if one inspects the EW(H$\alpha$) profile for GV galaxies in the highest mass bin (Fig. \ref{fig3}, top). In the central regions the median EW(H$\alpha$) approaches $\sim 1-2$ \AA, the characteristic value for LIER emission \citep{Belfiore2017}. The same regions show the lowest measured sSFR values ($\rm log(sSFR/yr^{-1})< -12.0$), which should, however, be interpreted with caution. Such low sSFRs are directly dependent on the LIER correction implemented by equation \ref{eq_correct}. In particular, assuming a lower value for ([SII]/H$\rm \alpha)_L$ causes more regions to appear fully LIER-dominated ($\rm f_{SF}=0$), and therefore further decreases the inferred sSFR in these regions. We note that the systematic effect of this correction is largest in the central regions of GV galaxies, but does not affect the derived profiles in the SFMS, except for the central regions of the most massive bin, which includes some cLIER galaxies.

\begin{figure*} 	
	\includegraphics[width=1.0\textwidth, trim=45 20 60 0, clip]{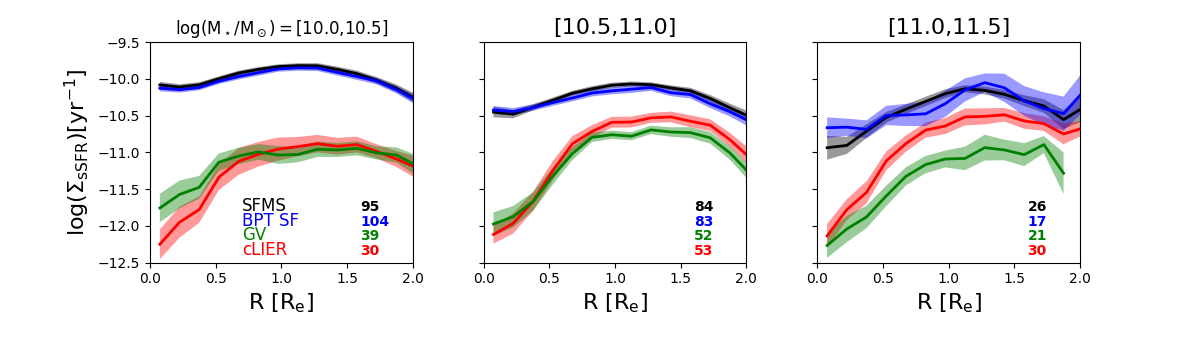}
	\caption{The stacked sSFR radial profiles for different categories of galaxies: SFMS, BPT star forming, green valley and cLIER. The number of galaxies in each stack is shown in each panel. The SFMS and GV stacks are the same as in Fig. \ref{fig3}, bottom.} 
	\label{fig5}
\end{figure*}

In Fig. \ref{fig5} we compare the sSFR profiles of SFMS and GV galaxies together with those of BPT-classified star forming and cLIER galaxies. We only consider mass bins with $\rm log(M_\star/M_\odot)> 10.0$, where we have sufficient statistics for cLIERs. There is a close correspondence between the GV galaxies and cLIERs on one hand, and between the SFMS and BPT-classified star forming galaxies on the other. In particular, most cLIERs lie in the GV except for the highest-mass bin, where some cLIERs fall on the SFMS. Overall cLIERs make up 56\% of the galaxies in the GV considered in this work. On the other hand, BPT-classified star forming galaxies make up 92\% of the SFMS. 
As can be appreciated from Fig. \ref{fig5} the presence of central LIER emission is an excellent predictor of sSFR suppression, even at large galactocentric radii. This points to central LIER emission as a tell-tale sign of the quenching process. 


\begin{figure*} 	
	\includegraphics[width=1.0\textwidth, trim=20 20 60 0, clip]{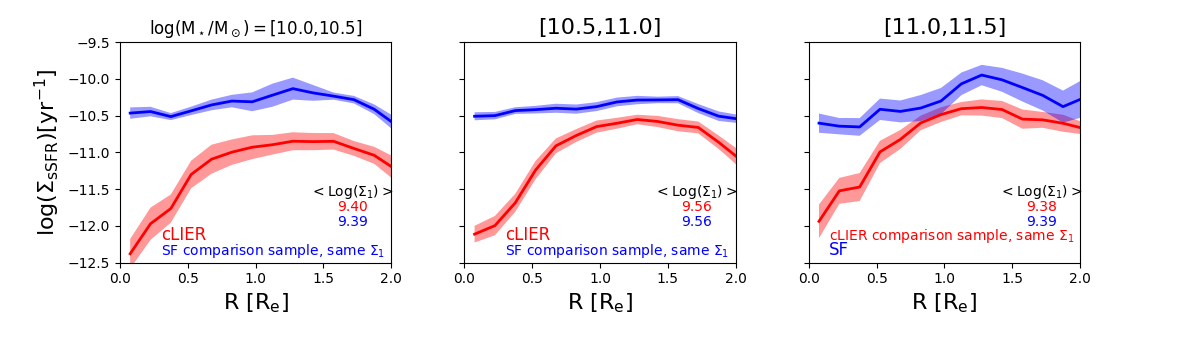}
	\caption{Same as Fig. \protect\ref{fig5}, but comparing the cLIER sample with a sample of star forming galaxies matched in mass and $\rm \Sigma_{1 \ kpc}$ (the stellar mass surface density within 1 kpc, as defined in \citealt{Fang2013}). The comparison demonstrates that cLIER galaxies have lower sSFR in their discs even when compared to a sample of star forming galaxies with comparable total mass and central mass surface density ($\rm \Sigma_{1 \ kpc}$).} 
	\label{fig6}
\end{figure*}

\subsection{Are bulges responsible for the sSFR suppression?}
\label{sec3.3}

In \cite{Belfiore2017} we showed that cLIER galaxies have higher concentration and S\'ersic indices than star forming galaxies of the same stellar mass. It is therefore worth investigating whether the sSFR difference between star forming and cLIER galaxies persists when considering galaxies of the same mass and central concentration. To make such a comparison we considered different observational proxies for bulge prominence, including the S\'ersic index (measured from $r$-band SDSS photometry), the concentration parameter ($\rm C= R_{90}/R_{50}$, where R are Petrosian radii enclosing respectively 90\% and 50\% of the $r$-band light) and the mean stellar mass surface density within 1 kpc from the galactic centre ($\rm \Sigma_{1 \ kpc}$, following the definition in \citealt{Fang2013}, measured directly from the MaNGA stellar mass maps). The choice of a 1 kpc aperture is motivated by the use of this physical scale in previous work, even though 1 kpc does not reflect the physical size of any well-defined galactic component. We note that 1 kpc corresponds to $\sim$ 0.2 $\rm R_e$ for galaxies of $\rm log(M_\star/M_\odot) \sim 10.25$ and $\sim$ 0.12 $\rm R_e$ for galaxies of $\rm log(M_\star/M_\odot) \sim 11.25$.

For each cLIER galaxy, we associate a star forming galaxy within the same stellar mass bin and with similar bulge proxy. We then compare the sSFR radial profiles of cLIER and the selected comparison sample of star forming galaxies in the different mass bins. We note that the highest mass bin ($\rm log(M_\star/M_\odot)=11.0-11.5$) contains more cLIER than star forming galaxies, so we reverse the procedure and find a comparison sample of cLIER galaxies matched in mass and bulge proxy to the few star forming galaxies present at these masses.

The results obtained by matching in concentration parameter, S\'ersic index and $\rm \Sigma_{1 \ kpc}$ are qualitatively similar, so we only show the results obtained matching by $\rm \Sigma_{1 \ kpc}$ (Fig. \ref{fig6}). The average $\rm \Sigma_{1 \ kpc}$ values for the two samples in the three considered mass bins are also shown in figure  \ref{fig6}, demonstrating the efficacy of our matching procedure. This analysis shows that at fixed mass and $\rm \Sigma_{1 \ kpc}$ there exist galaxies with both star forming and suppressed (cLIER) central regions. In other words, while cLIERs have higher central concentration at fixed mass  on average, a high $\rm \Sigma_{1 \ kpc}$ is a necessary but not sufficient condition for central quiescence. We conclude that a high central density cannot be the only cause for the central sSFR suppression observed in cLIERs and in the GV in general.


\section{Discussion}
\label{dis}

\subsection{The sSFR of star forming galaxies}
\label{dis.2}

\subsubsection{Inside-out growth and the role of the bulge}

The decrease in the EW(H$\alpha$) and sSFR profiles observed in the central regions of SFMS galaxies can be attributed to the combined effects of radially-dependent disc formation timescales and bulge assembly. According to the expectations from growth of structure in the cosmological context, the central regions of galactic discs are formed at earlier times (`inside-out' growth), and are therefore more evolved, less gas rich and will show lower sSFR than the outer disc, which is still un-evolved and gas rich \citep{Munoz-Mateos2011, Pezzulli2015a, Ibarra-Medel2016}. The inside-out growth paradigm is supported by extensive observational evidence, including disc colour gradients \citep{deJong1996, Bell2000, Munoz-Mateos2007}, studies of the stellar populations fossil record \citep{Sanchez-Blazquez2014, Delgado2015, Goddard2017b, Delgado2016} and gas phase metallicity gradients \citep{Boissier1999, Chiappini2001, Molla2005}. Moreover \cite{Lilly2016a} show that implementing inside-out growth in a fashion which is consistent with the redshift evolution of the sSFR and the size evolution of star forming galaxies naturally leads to sSFR gradients, with central regions of galaxies having lower sSFR simply because of their earlier formation timescale.

The importance of the bulge on the integrated sSFR of galaxies has been subject of much recent work. \cite{Abramson2014}, for example, argue that bulges should be considered passive components of galaxies, contributing mass but negligible SFR. Under this assumption, the observed sub-linear slope of the SFMS, especially at the high-mass end, can be explained by an increase in bulge mass. This interpretation is consistent with recent work showing that the slope of the SFMS depends on stellar mass, with the relation being closer to linear for lower-mass galaxies \citep{Gavazzi2015}.
Other authors, however, have found that bulges are an important source of scatter on the SFMS, but even after their removal the SFMS main sequence remains sub-linear at the high-mass end, both in SDSS \citep{Guo2015a} and at higher redshifts \citep{Whitaker2015, Schreiber2016}. This implies a mass-dependence in the sSFR of discs on the SFMS, which should be observable in spatially-resolved data. 

A notably different approach to this question is taken in the recent work by \cite{Catalan-Torrecilla2017}, who infer the sSFR of disc and bulges using IFS data from the CALIFA survey. These authors depart from the assumption that bulges host no star formation, and assume instead that discs and bulges share the same spectrum at each position on sky. They then make use of photometric bulge-disc decompositions to assign a fraction of the flux at each location to bulge and disc respectively. Their bulge sSFR are therefore best interpreted as the sSFR of the area of the galaxy dominated by bulge light. This definition, however, means that their derived disc sSFR are underestimated with respect to the \cite{Abramson2014} definition, since some of the total SFR is attributed to the bulge component. Despite their different approach \cite{Catalan-Torrecilla2017} also find that disc sSFR decreases with stellar mass, going from $\sim ~ 10^{-9.8}$ for $\rm log(M_\star/M_\odot)=9.25$ to $\sim 10^{-11.0}$ for $\rm log(M_\star/M_\odot)=11.25$, in good agreement with the results presented here.

\begin{figure*} 	
	\includegraphics[width=0.65\textwidth, trim=0 0 0 0, clip]{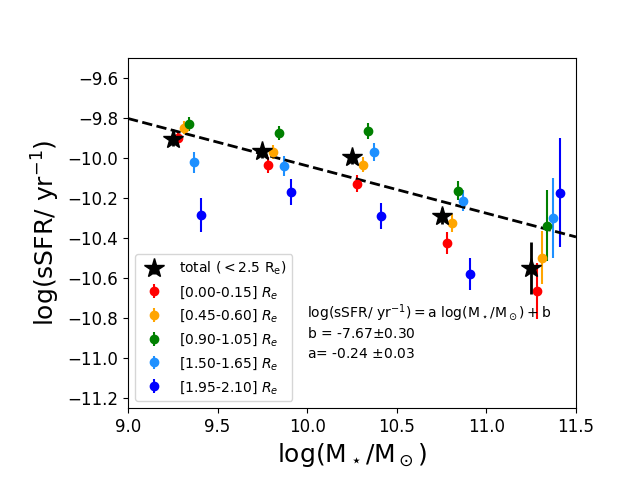}
	\caption{The sSFR of BPT-classified star forming galaxies in different mass bins, both integrated within $\rm 2.5 \ R_e$ (black stars) and in different radial bins derived from the stacked profiles (see Fig. \ref{fig5}). Different radial bins are colour-coded as described in the legend and are shown with slight offsets in the x direction for ease of visualisation. The black dashed line is the linear best-fit relation to the median integrated sSFR values. The parameters of this best-fit relation are shown in the bottom right corner.} 
	\label{fig7}
\end{figure*}

\begin{table*}
	\caption{log(sSFR/$\rm yr^{-1}$) for BPT-classified star forming galaxies in different stellar mass bins, both integrated over a $\rm 2.5 \ R_e$ aperture and in some representative radial bins. The mean redshift for galaxies in each mass bin is also noted. The log(sSFR/$\rm yr^{-1}$) data is plotted in Fig. \ref{fig7}.}
	\label{table2}
	\centering
	\begin{tabular}{ l c c c c c }
		     & $\rm log (M_\star/M_\odot)$ & $\rm log (M_\star/M_\odot)$ & $\rm log (M_\star/M_\odot)$ & $\rm log (M_\star/M_\odot)$ & $\rm log (M_\star/M_\odot)$ \\
		     & [9.0-9.5] &[9.5-10.0] & [10.0-10.5]& [10.5-11.0]& [11.0-11.5]  \\
	
		\hline
		integrated log(sSFR) ($\rm < 2.5 \ R_e$) & -9.91 $\pm$ 0.03  &  -9.97 $\pm$ 0.03 & -10.00 $\pm$ 0.04   &  -10.29 $\pm$ 0.04  &  -10.55 $\pm$ 0.13 \\
		\hline
		log(sSFR) ($\rm [0.00-0.15] \ R_e$) & -9.90 $\pm$ 0.03  &  -10.04 $\pm$ 0.04 & -10.13 $\pm$ 0.04   &  -10.43 $\pm$ 0.05  &  -10.67 $\pm$ 0.14 \\
		log(sSFR) ($\rm [0.45-0.60] \ R_e$) & -9.85 $\pm$ 0.03  &  -9.97 $\pm$ 0.03 & -10.03 $\pm$ 0.04   &  -10.33 $\pm$ 0.04  &  -10.50 $\pm$ 0.13 \\
		log(sSFR) ($\rm [0.90-1.05] \ R_e$) & -9.83 $\pm$ 0.03  &  -9.88 $\pm$ 0.03 & -9.87 $\pm$ 0.04   &  -10.16 $\pm$ 0.05  &  -10.34 $\pm$ 0.18 \\
		log(sSFR) ($\rm [1.50-1.65] \ R_e$) & -10.02 $\pm$ 0.05  &  -10.04 $\pm$ 0.05 & -9.97 $\pm$ 0.05   &  -10.22 $\pm$ 0.05  &  -10.3 $\pm$ 0.2 \\
		log(sSFR) ($\rm [1.95-2.10] \ R_e$) & -10.29 $\pm$ 0.09  &  -10.17 $\pm$ 0.07 & -10.29 $\pm$ 0.06   &  -10.58 $\pm$ 0.08  &  -10.2 $\pm$ 0.3 \\
		\hline
		mean redshift $<z>$  &  0.028 & 0.032  & 0.039 & 0.052 &  0.088  \\
		mean (Petrosian Elliptical) $\rm R_e$ [kpc]  &  3.0 & 3.6  & 4.9 & 6.0 &  8.9  \\
		\hline
	\end{tabular}
\end{table*}

In this work we do not tackle the issue of bulge-disc decompositions, but we study the shape of radial sSFR of galaxies of different masses. In Fig. \ref{fig7} and Table \ref{table2} we present a collection of the results on the sSFR of BPT-classified star forming galaxies, considering both their integrated sSFR values (as derived in Sec. \ref{sec2.3.1}) and their radial profiles. In Sec. \ref{sec2.3.1} we show that star forming galaxies show a decreasing sSFR as a function of $\rm M_\star$, when both quantities are integrated within a $\rm 2.5 \ R_e $ aperture. By fitting this trend with a linear relation we obtain a slope of -0.24 $\pm$ 0.03 dex. The stellar mass dependence of the sSFR is present in different radial bins, although for the largest radii $\rm \sim 2.0 \ R_e$ the trend is less well defined. At $\rm \sim 1.5 \ R_e$ the difference in log(sSFR) between the least and most massive bin is still significant and amounts to 0.28 dex.

In order to obtain a rough estimate of the potential effect of the bulge on our measurements of sSFR at large radii, we have matched the sample of galaxies used in this work with the \cite{Simard2011} set of bulge-disc decompositions, based on SDSS photometry. For the 522 galaxies in common with the Simard catalogue, we find that the median bulge effective radius is $\rm \sim 0.5 \ R_e$ (where $\rm R_e$ refers to the effective radius of the galaxy as a whole), with no strong dependence on stellar mass. Moreover, we find that the disc is the dominant light component at $\rm \sim 1.5 \ R_e$ for $\sim 95\% $ of the galaxies in our sample. Our data, therefore, does not support the hypothesis that discs of galaxies of different stellar masses grow at constant sSFR.
 
While measurements at large radii allow us to directly probe the disc component free from the confounding effect of the bulge, it is worth remembering that the outer disc is of limited importance in setting the total SFR of the galaxy. Using the curve of growth analysis of Appendix \ref{app1}, we infer that the radii that contribute most to the total SFR are  $\rm 0.5 < R/R_e <1.0$ and the SFR contributed by $\rm R> 1.5\ R_e$ is only 9\% of the total SFR within $\rm 2.5\ R_e$. 

\subsubsection{The SFMS in the local and high-redshift Universe}

In \cite{Spindler2018} we show that sSFR profiles for individual galaxies are broadly bimodal: either flat or centrally suppressed. As we have shown in this work, centrally suppressed profiles are mostly associated with the GV and central LIER emission. In \cite{Spindler2018} we further demonstrate that the occurrence of central suppression is a function of mass, but not of environment. This fact suggests that central suppression should be associated with processes internal to the galaxy and its halo.

Other works have also probed sSFR gradients in local galaxies, mostly via proxies like the $NUV-r$ colour \citep{Pan2016b, Bouquin2017, Lian2017}, generally finding negative gradients. \cite{Pan2016b}, for example, find that both inner and outer regions of galaxies show increasingly red $NUV-r$ colour with increasing mass. 
More recently, similar studies of UV-optical colour gradients have become possible at higher redshift, making use of the exquisite photometric data from CANDELS \citep{Liu2016a, Wang2017b}.
\cite{Wang2017b}, for example, make use of $UVI$ colour profiles to derive dust attenuation and sSFR for galaxies of $\rm log(M_\star/M_\odot)=9.0-11.0$ in the redshift range $0.4 < z < 1.0$. They find nearly flat sSFR profiles over the whole mass and redshift range, except for galaxies with $\rm log(M_\star/M_\odot)=10.5-11.0$, which are consistent with having a $\sim$ 0.1 dex decrease in sSFR in their central regions. These observations imply that the centres of massive galaxies develop the more substantial sSFR suppressions observed in this work between z=0.4 and z=0.

At $\rm z \sim 1$ \cite{Nelson2016} derive sSFR radial profiles from \textit{Hubble Space Telescope} grism observations of the H$\alpha$ line. Although in \cite{Nelson2016} the sSFR profiles for massive galaxies show a central dip, \cite{Wang2017b} show that this central decrease can be due entirely to the effect of dust attenuation gradients \citep{Nelson2016a}. At $\rm z \sim 2$ \cite{Tacchella2015} presents  deep adaptive-optics corrected SINFONI observations of the H$\alpha$ line in 22 galaxies, and find a central suppression in sSFR of $\sim$ 1 dex for galaxies of $\rm log(M_\star/M_\odot) \sim 11$. In a subsequent study \cite{Tacchella2017_arxiv} obtain rest-frame UV photometry for 10 galaxies in their sample and derive spatially resolved dust attenuation profiles. After correcting for dust extinction, they find nearly flat sSFR profiles for $\rm log(M_\star/M_\odot) =10.0 -11.0$ and confirm a central drop of $\sim$ 1 dex in sSFR in the centre of galaxies of $\rm log(M_\star/M_\odot) =11.0 -11.5$. Such massive galaxies at z $\sim$ 2 are likely progenitors of passive ellipticals at z=0 and are therefore not directly related in an evolutionary sense to the MaNGA galaxies studied in this work.

\subsection{sSFR suppression in the green valley}
\label{dis.3}

\subsubsection{Central quiescence and bulges in GV galaxies}
Based on studies of their S\'ersic indices and stellar mass profiles, GV galaxies are known to host larger bulges than blue cloud galaxies of the same stellar mass \citep{Schiminovich2007, Fang2013}. Hence their sSFR profiles are expected to be more centrally suppressed than those of blue cloud galaxies of same total mass. In this work we have additionally shown that the sSFR profiles of GV galaxies differ from those of the SFMS galaxies also in their discs (i.e. at large galactocentric radii). The implication is that the suppressed integrated sSFR of GV galaxies cannot be reproduced simply by adding more mass in the quiescent bulge, but a change in the properties of the disc is also necessary. The findings from Sec. \ref{sec3.3} further confirm that having a high central concentration is not a sufficient condition for central quiescence. Bulges are therefore not uniquely responsible for the deviation of cLIER galaxies from the SFMS.

Similar conclusions are reached by studying the colours of discs and bulges of GV galaxies in the nearby Universe \citep{Vulcani2015, Morselli2017}. In \cite{Vulcani2015}, for example, it is found that green late type galaxies have slightly larger bulges than blue galaxies of the same mass, but both disc and bulges of the green systems are experiencing reduced levels of star formation.

Comparing to higher redshift work is challenging, but our results here echo those of \cite{Nelson2016}, who conclude that at $\rm z \sim 1$ the sSFR of galaxies lying below the SFMS is already suppressed with respect to galaxies on the SFMS, at all radii. More recently \cite{Tacchella2017_arxiv} also find a higher degree of sSFR suppression for galaxies below the SFMS at $\rm z \sim 2$.

\subsubsection{A critique of the `quenching' paradigm: a quasi-static GV}

Models are able to reproduce the tight nature of the SFMS by assuming galaxies live in an equilibrium between cosmological accretion rate, star formation and outflow rate \citep{White1991, Bouche2010, Lilly2013, Peng2014}. Galaxies below the SFMS, lying in the GV, are generally assumed to have been perturbed off this equilibrium, and in the process of transitioning towards the red sequence. Galaxies transiting in the opposite direction (from the red sequence to the blue cloud) as a consequence of a `rejuvenation' event are found by models to be a sub-dominant population in the GV \citep{Trayford2016, Pandya2017a}.

Following a somewhat different approach, quenching can also be defined as the \textit{smooth} crossing of galaxies to arbitrarily low sSFR \citep{Gladders2013, Vulcani2015, Abramson2016, Eales2017}. While this definition explicitly neglects very fast quenching events, it has the advantage of regarding quenching as a more `natural' endpoint of galaxy evolution, which may not necessarily require a triggering event. 

In either case, the timescale associated with crossing the GV remains a point of contention. This timescale was originally thought to be very short, in order to explain the lack of a substantial population of galaxies at intermediate optical colours \citep{Baldry2004}. As shown in \cite{Wyder2007} and \cite{Salim2007}, the use of UV-optical colours highlights the presence of a substantial population of galaxies at intermediate sSFR, which was previously not evident in the optical colour-magnitude diagram. Exploiting the $NUV-r$ selection criterion, \cite{Martin2007} derived timescales of the order of $\sim$ Gyr for galaxies to cross the GV. Intermediate redshift data suggests that this timescale should decrease at higher redshift \citep{Goncalves2012}. 

\cite{Schawinski2009} compared UV-optical colours of $\rm z \sim 0$ GV galaxies with stellar population models to argue that, while early-type galaxies may have transitioned quickly trough the GV (see also \citealt{Schawinski2007}), late type galaxies are likely to be on a slow pathway to quiescence. They suggest `strangulation' (i.e. the reduction of the cold gas supply from the cosmological accretion) as a viable mechanism for the slow quenching mode. \cite{Smethurst2015} used UV-optical colours in a Bayesian framework to fit two-parameter star formation histories and argue in favour of the existence of both a fast and a slow quenching channel through the GV. \cite{Eales2017} make use of 21 band photometry, including \textit{Herschel} far-IR data, to trace the sSFR in a volume-limited sample of galaxies and similarly argue for the need of a slow quenching process at $\rm z=0$ to reproduce the significant number of objects observed at intermediate sSFR, especially at high masses.

\cite{Abramson2016} use an ensemble of log-normal star formation histories, matching the overall star formation history of the Universe, to argue that galaxies which appear quenched today simply had more compressed star formation histories, and have therefore already exhausted their cosmic  supply of gas (although they also acknowledge the need to maintain these galaxies quiescent after they gas supply has been exhausted, possibly via AGN heating). This simple model naturally predicts that galaxies crossing the GV at higher redshift had more compressed star formation histories and were thus transitioning on faster timescales (`fast quenching'). Galaxies crossing the GV today, on the other hand, correspond to more extended star formation histories and are therefore exhausting their gas on slower timescales. 
 
The existence of widespread star formation in the discs of GV galaxies with quiescent central regions is now well established, thanks to UV observations first \citep{Kauffmann2007, Thilker2007, Salim2010, Salim2012, Fang2012}, and large IFS surveys later \citep{Belfiore2017}. Moreover, in \cite{Belfiore2017} we have demonstrated the absence of any significant kinematic misalignment between gas and stars in cLIER galaxies, which show regular disc kinematics (this conclusion is to be contrasted with the kinematics of extended LIER galaxies, which are found to be strongly misaligned). Further insight can be obtained from a more detailed study of the stellar population ages in the GV. Although this is beyond the scope of this work, we point out that the central regions of cLIER galaxies show uniformly old stellar populations \citep{Belfiore2017}. This evidence suggests that neither fast gas expulsion nor a recent major merger can easily explain the star formation in the green valley, and instead supports the idea of the GV as a `quasi-static' population subject to a slow quenching process (borrowing the terminology from \citealt{Salim2014}). 

This slow quenching would maintain the observed disc morphology while building up the bulge, possibly via interactions and minor mergers \citep{Bundy2010}, or bar-driven secular evolution \citep{Gavazzi2015}. In this scenario, GV galaxies, once fully quenched, may give rise to the population of passive discs seen both locally \citep{Masters2010, Morselli2017} and at high redshift \citep{Bundy2010}. Our observations, however, do not preclude the existence of a fast quenching channel, exemplified, for example, by the population of post-starburst galaxies \citep{Wild2017}. Since these galaxies would transition through the GV on fast timescales, the majority of galaxies currently observed in the GV would still belong to the quasi-static population.

\subsubsection{The physics of the slow quenching mode}


In light of the need for a slow quenching mechanism two main classes of processes can be responsible for the observed reduction in sSFR in GV discs: `preventive' processes, leading to a reduction in the gas content of GV galaxies, or `sterilising' processes, predicating a reduction in the star formation efficiency of the cold gas. Purely `ejective' processes, which would require the expulsion of the gas and are unlikely to lead to slow quenching, may still act in combination with other (e.g. preventive) processes to lead to the observed quenching signature.

AGN feedback, for example, is not only ejective but can also be preventive in nature, and lead to reduced rates of gas accretion onto GV galaxies, causing a slow fading of their discs and decrease in their overall sSFR.
More generally, any process that prevents cold gas accretion from the IGM from reaching the galaxy will lead to slow quenching over the entire galaxy disc (this process is sometimes referred to as `strangulation', \citealt{Peng2015}). 

The role of AGN in the GV has long been debated in the literature \citep{Sanchez2004, Nandra2007, Schawinski2014, Trump2015}. Making use of MaNGA data, \cite{Sanchez2017} present the most detailed study of the relation between AGN and the GV to date. They conclude that optically-selected strong AGN lie preferentially in the GV. They also find that the sSFR profiles of AGN and non-AGN galaxies in the GV are qualitatively similar. It remains difficult, however, to relate the short timescale associated with the AGN duty cycle with the longer timescale associated with the quenching process, and therefore determine whether the AGN is the cause, or a concurrent phenomenon, to the observed sSFR suppression.

 Alternatively a mechanism is required to make star formation less efficient in bulge-dominated galaxies \citep{Martig2009}  or galaxies which host bars \citep{Cheung2013, Emsellem2015a, Gavazzi2015}. It is likely that both processes are at play, since galaxies below the SFMS are found to have both lower gas fractions and lower star formation efficiencies \citep{Saintonge2012, Genzel2015, Saintonge2016}. A pilot study of resolution-matched ALMA and MaNGA data \citep{Lin2017c} in three GV galaxies confirms this scenario, although larger samples of resolved molecular gas observations would prove useful in understanding the relative changes in star formation efficiency and gas fraction between the bulge and disc of GV galaxies.

Finally, the required quenching process does not act as a step function in radius, only affecting the innermost regions and gradually moving outwards, but has an effect on the entire galaxy. This finding echoes that of \cite{Lian2017}, who considers both the $NUV - u$ and $u-i$ colours, to gain insight into the timescales of inside out growth and quenching in star forming galaxies. Their work favours a model where, after an initial period of inside-out growth, inner and outer regions of the galaxy `quench' synchronously. It is worth noting, however, that \cite{Lian2017} define quenching as a period where the SFR is declining exponentially with a faster $e$-folding time than during the growth phase. In this work, however, we have limited ourselves to studying the ratio between current star formation (SFR) and its time integral ($\rm M_\star$). We are therefore unable to discriminate between a slowdown of `growth' and discontinuity in the star formation history of the galaxy, which could be referred to as `quenching'.

Even when star formation histories are explicitly derived \citep{Vulcani2015, Ibarra-Medel2016}, it is unclear whether the finite age resolution afforded by such archaeological methods will be able to meaningfully discriminate between growth and quenching on timescales longer than those giving rise to classical post-starburst features. Full spectral fitting is most accurate at reconstructing the recent star formation history, but the age resolution quickly worsens for older populations, making it very challenging to determine discontinuities in the star formation history. Comparing the stellar population properties of $\rm z=0$ galaxies with those of their higher-redshift progenitors may represent a way to gain more insight into the quenching timescale. Current deep spectroscopic surveys (e.g. LEGA-C, \citealt{VanderWel2016a}) and the spectroscopic capabilities of the \textit{James Webb Space Telescope} will soon provide suitable datasets to address this question.

\section{Conclusions}
\label{concl}

We derive sSFR and EW(H$\alpha$) radial profiles in a representative sample of nearby ($<z> \sim 0.03$) galaxies with resolved spectroscopy from SDSS-IV MaNGA in order to study spatially resolved star formation on the star formation main sequence and in the green valley. We derive integrated SFR and $\rm M_\star$ from the MaNGA data and obtain a sub-linear star formation main sequence (with a slope of 0.73 and scatter of 0.39 dex). We define the green valley with respect to the star formation main sequence as the locus in the $\rm M_\star$-SFR plane more than 1$\sigma$ below the relation.

The main goal of this work is to elucidate the role of galactic subcomponents (the bulge and the disc) in setting the sSFR of galaxies, both on the star formation main sequence and the green valley. Central LIER (cLIER) galaxies, which are defined to have quiescent, LIER central regions and star forming outer regions, are a particularly interesting class of objects in this context, as they live preferentially in the GV and the upper-mass end of the main sequence. Below we summarise the main findings of this work.

\begin{enumerate}
	\item{The sSFR and EW(H$\alpha$) profiles on the main sequence (and in BPT-classified star forming galaxies) change smoothly as a function of total stellar mass. While low-mass galaxies show mostly flat profiles, with increasing mass we find larger central sSFR suppressions. This trend may reflect the increasing importance of the bulge component, but also the natural evolution of the sSFR in the context of inside-out growth, where central regions are more evolved and therefore less gas rich. At large radii ($\rm R > 1.5 \ R_e$) the sSFR of galaxies of different masses shows a gradual decline and are in closer agreement with each other than at small radii. Even at $\rm R > 1.5 \ R_e$, however, we find that lower-mass galaxies have higher sSFR than high-mass ones (with 0.28 dex sSFR difference between $\rm log(M_\star/M_\odot) =[ 9.0-9.5]$ and $\rm log(M_\star/M_\odot) =[11.0-11.5]$). This demonstrates that sSFR is mass dependent even in regions dominated by the disc component.  }
	
	\item{The sSFR and EW(H$\alpha$) profiles of green valley galaxies show a suppression with respect to mass-matched main sequence galaxies at all radii. The fact that this suppression persists out to the largest radii probed in this work (i.e. $\rm 2 \ R_e$) indicates that the star formation properties of the discs of green valley galaxies are fundamentally different than those of the discs of main sequence galaxies of the same mass. }
	
	\item{We use the [SII]/H$\alpha$ line ratio to estimate the amount of star formation that can be present in BPT-classified LIER regions. This procedure assumes that a faction of the observed H$\alpha$ flux in LIERs may be due to residual star formation. While the effect of LIERs is negligible for integrated SFRs and on the main sequence, it is of fundamental importance in the green valley. In fact, LIER spaxels are increasingly common in the central regions of green valley galaxies with $\rm log(M_\star/M_\odot) > 10.0$. In agreement with the general green valley population, cLIER galaxies show suppressed sSFR with respect to the SFMS not only in their central regions, but also at large galactocentric distances.}
	
	\item{While cLIER galaxies present higher central mass concentrations on average, when controlling for both $\rm M_\star$ and $\rm \Sigma_{1~ kpc}$ (the stellar mass density within 1 kpc), cLIERs still show lower sSFR than a matched sample of star forming galaxies. This demonstrates that the bulge cannot be the only driver of central quiescence.}

\end{enumerate}

These observations demonstrate that the GV is a complex space. While GV galaxies have larger bulges than SFMS galaxies at fixed mass, they also have fundamentally different discs, with suppressed sSFR. These observations support a view of the GV as a quasi-static population, requiring a slow quenching process, uniformly affecting the entire galaxy.


\section*{Acknowledgements}
\begin{footnotesize}
	We thank the anonymous referee for a very supportive and thought-provoking report. F.B. thanks Alvio Renzini for his suggestion to further quantify the role of central LIER galaxies on the star formation main sequence. F.B. also thanks the organizers and participants of the Lorentz Workshop `The physics of quenching massive galaxies at high redshift' for the many fruitful discussions.
	F.B. and R.M. acknowledge funding from the Science and Technology Facilities Council (STFC). R.M. acknowledges funding from the European Research Council (ERC), Advanced Grant 695671 `QUENCH'. M.B. acknowledges funding from NSF/AST-1517006.
	This work makes use of data from SDSS-IV. Funding for SDSS has been provided by the Alfred P.~Sloan Foundation and Participating Institutions. Additional funding towards SDSS-IV has been provided by the U.S. Department of Energy Office of Science. SDSS-IV acknowledges support and resources from the Centre for High-Performance Computing at the University of Utah. The SDSS web site is {\tt www.sdss.org}. This research made use of Marvin, a core Python package and web framework for MaNGA data, developed by Brian Cherinka, Jos\'e S\'anchez-Gallego, and Brett Andrews \citep{Cherinka2017}.
	SDSS-IV is managed by the Astrophysical Research Consortium for the Participating Institutions of the SDSS Collaboration including the  Brazilian Participation Group, the Carnegie Institution for Science, Carnegie Mellon University, the Chilean Participation Group, the French Participation Group, Harvard-Smithsonian Center for Astrophysics, Instituto de Astrof\'isica de Canarias, The Johns Hopkins University, Kavli Institute for the Physics and Mathematics of the Universe (IPMU) / University of Tokyo, Lawrence Berkeley National Laboratory, Leibniz Institut f\"ur Astrophysik Potsdam (AIP),  Max-Planck-Institut f\"ur Astronomie (MPIA Heidelberg), Max-Planck-Institut f\"ur Astrophysik (MPA Garching), Max-Planck-Institut f\"ur Extraterrestrische Physik (MPE), National Astronomical Observatory of China, New Mexico State University, New York University, University of Notre Dame, Observat\'ario Nacional / MCTI, The Ohio State University, Pennsylvania State University, Shanghai Astronomical Observatory, United Kingdom Participation Group, Universidad Nacional Aut\'onoma de M\'exico, University of Arizona, University of Colorado Boulder, University of Oxford, University of Portsmouth, University of Utah, University of Virginia, University of Washington, University of Wisconsin, Vanderbilt University, and Yale University.
	
	The MaNGA data used in this work is publicly available at {\tt http://www.sdss.org/dr13/manga/manga-data/}.

\end{footnotesize}	

\bibliography{Feb2018}
\bibliographystyle{mnras}

%
%
\appendix 

\section{Defining the green valley}
\label{app1}

In this appendix we study the relation between different definitions of the GV. In particular, we show that the $NUV-r$ GV corresponds closely to the GV as defined in this work, while the GV defined in optical colours is heavily contaminated by blue cloud star forming galaxies.

\begin{figure} 
	
	\includegraphics[width=0.52\textwidth, trim=10 0 0 20, clip]{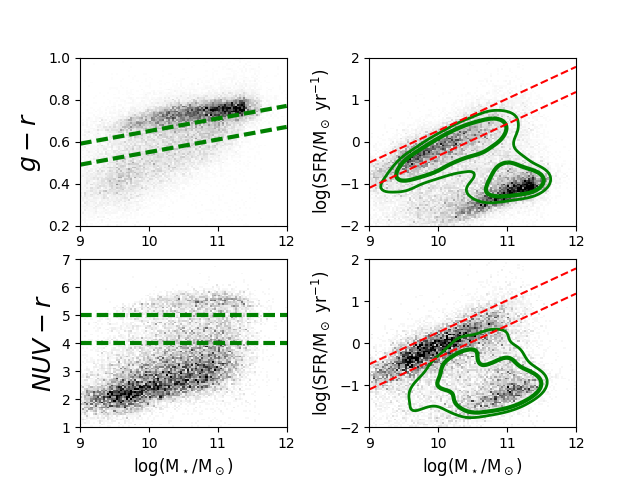}
	\caption{\textit{Top left}: The $g-r$ rest-frame colour - mass diagram for the MaNGA parent sample. The definition of the GV in $g-r$ colours of \protect\cite{Lackner2012} is shown by the dashed green lines. \textit{Top right}: The SFR-$\rm M_\star$ diagram for the MaNGA parent sample, showing the position of the galaxies selected using the $g-r$-GV cuts of \protect\cite{Mendel2013} using green contours, enclosing respectively 50\% and 70\% of the $g-r$-GV galaxies. The plot demonstrates that the adopted cut in $g-r$ select a large number of galaxies on the SFMS, while also selecting red sequence objects. Overall the colour cut does not preferentially select galaxies between the SFMS and the upper limits representing passive galaxies.
	\textit{Bottom}: Same as top panels, but using $NUV-r$ rest-frame colours instead of $g-r$. The red sequence appears less populated than using $g-r$ colours because some red sequence galaxies are undetected in GALEX. The GV is defined to have $4<NUV-r<5$. This selection criteria leads to a sample of galaxies located preferentially below the SFMS, as shown by the green contours in the bottom right panel. } 
	\label{figApp1}
\end{figure}

In order to test different definitions of the GV, we consider the MaNGA targeting catalogue, consisting of all SDSS galaxies meeting the MaNGA target selection criteria ($\sim$ 30000 galaxies). We extract elliptical Petrosian photometry in the \textit{GALEX} and SDSS bands from the extended NSA catalogue and stellar masses and SFR from the MPA-JHU catalogue \citep{Brinchmann2004}. We note that the photometry used here is k-corrected and corrected for Galactic extinction, but not corrected for attenuation intrinsic to the galaxy.
We define the GV in $NUV-r$ and $g-r$ using the following cuts respectively (1) $4<NUV-r<5$ and (2) $\rm  0.55+0.06 \ log(M_\star/M_\odot) < $ $g-r$ $\rm < 0.65+0.06 \ log(M_\star/M_\odot)$ \citep{Mendel2013}. Fig. \ref{figApp1}, left panels, shows the position of the GV defined in this way in the colour-mass diagram. 

We then consider the position of GV galaxies selected using these two different cuts in the $\rm M_\star$-SFR plane (Fig. \ref{figApp1}, right panels). The SFMS definition of \cite{Renzini2015} is adopted here (dashed red lines  in Fig. \ref{figApp1}, right panels, correspond to the SFMS $\pm 0.3$ dex), which is based on stellar masses and SFR from the MPA-JHU catalogue. 

We find that, using the $g-r$ definition of the GV, galaxies are preferentially selected either on the lower envelope of the SFMS or in the passive population. Very few intermediate sSFR galaxies are actually selected (green contours in Fig. \ref{figApp1}, right panels corresponds to the position of the GV-selected galaxies). The $g-r$ GV selection is therefore not only biased by the inclusion of star forming galaxies, but also avoids intermediate sSFR systems (the same conclusion is illustrated in \citealt{Salim2014}). Using the $NUV-r$ definition of the GV, on the other hand, we select preferentially galaxies below the SFMS. 

We quantify the contamination of star forming galaxies using these two definitions of the GV by calculating the fraction of GV galaxies that lie within 0.3 dex of the SFMS. Using the $g-r$ green valley definition, the contamination fraction is 30\%, while using the $NUV-r$ definition it decreases to 7\%. We conclude that $NUV-r$ selection is efficient at selecting intermediate sSFR systems, while one should avoid selecting GV galaxies using $g-r$ colours.


\section{Aperture corrections and comparison with literature stellar masses}
\label{app2}

\begin{figure} 
	\centering
	\includegraphics[width=0.45\textwidth, trim=0 0 0 0, clip]{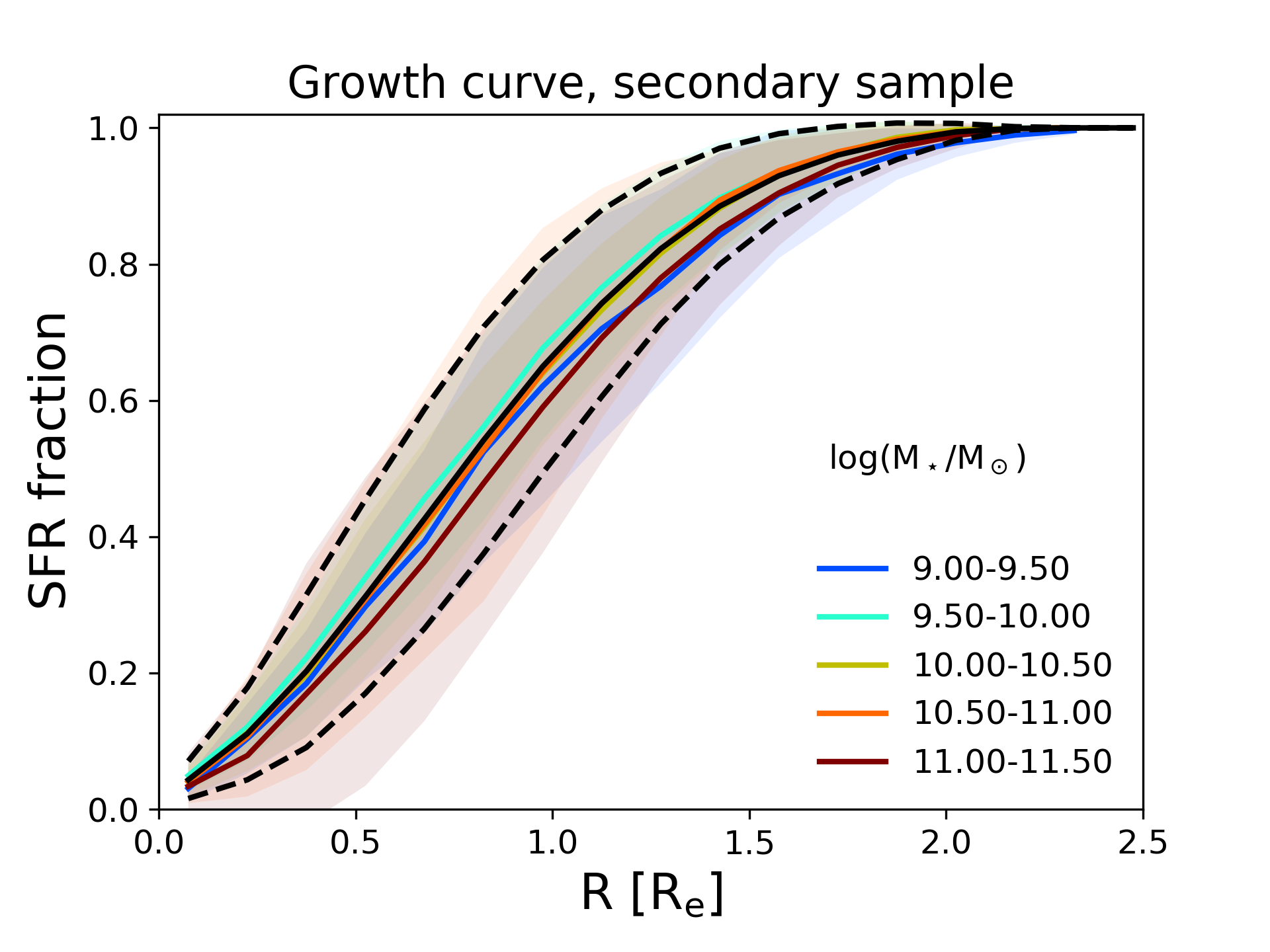}
	\includegraphics[width=0.45\textwidth, trim=0 0 0 0, clip]{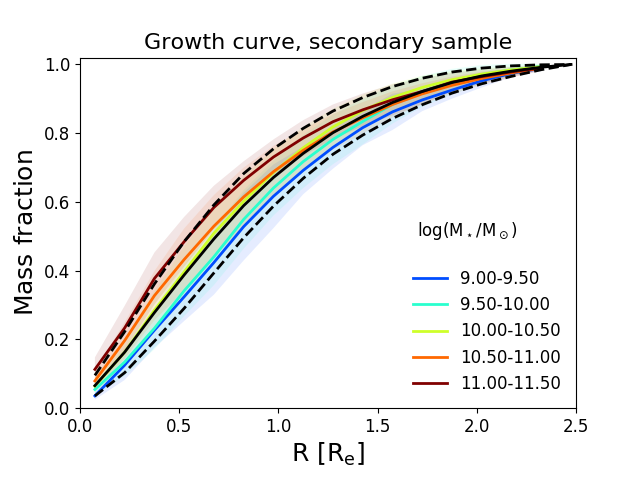}
	\caption{Growth curves for the SFR (top) and $M_\star$ (bottom) derive from the MaNGA data for the secondary sample galaxies. For each quantity the growth curve is define by the integral out to radius R normalised by the total within $\rm 2.5 \ R_e$. We show the growth curves for galaxies of different total stellar masses (here we use stellar masses derive from S\'ersic photometry from the NSA catalogue). The curves for different profiles are shown in colour, as outlined in the legend. The median profile and its standard deviation is shown with black solid and dashed curves.} 
	\label{figApp2.1}
\end{figure}

In this appendix we discuss the strategy followed in this paper to compute aperture corrections for the SFR and $\rm M_\star$ derived from the MaNGA data. We then compare the derived stellar masses with previous estimates from the literature based on fits to the integrated photometry and comment on the effect on the slope of the SFMS.

In order to derive an aperture correction, we have used the secondary sample to compute the cumulative SFR and $\rm M_\star$ within a given deprojected radial distance from the galaxy centre in units of $\rm R_e$, as in a classical growth curve analysis. In Fig. \ref{figApp2.1} we show cumulative profiles normalised to $\rm 2.5 \ R_e$. Profiles are shown for galaxies of different stellar masses (here the NSA mass derived from S\'ersic photometry) as detailed in the legend, while the median profile and its standard deviation are shown with black solid and dashed lines respectively. The SFR growth curves shows a larger scatter than the stellar mass one, reflecting larger differences in individual SFR profiles, but almost no dependence on stellar mass, demonstrating that all SFR profiles have a similar shape (close to an exponential decline). Considering the median profile for the whole sample an aperture extending to $\rm 1.5 \ R_e$ contains 91\% of total SFR within $\rm 2.5 \ R_e$. 

The stellar mass growth curves, on the other hand, show a systematic variation with total mass of the galaxy. The trend is consistent with more massive galaxies being more concentrated and therefore reaching a higher fraction of their integrated mass at smaller radii. This mass-dependent trend, however, only minimally affects our conclusions for the radial range $\rm 1.5-2.5 \ R_e$, where the growth curves for different profiles are the same within a few percent. For the median stellar mass growth curve, and $\rm 1.5 \ R_e$ aperture contains 87\% of the mass within $\rm 2.5 \ R_e$. Finally, we fit fifth order polynomials to the median SFR and $\rm M_\star$ growth curves and use these functions to compute aperture corrections. The best fit polynomial coefficients are shown in Table \ref{table_app_corr}.

\begin{table}
	\caption{Coefficients for the fifth-order polynomial fits to the SFR and $\rm M_\star$ median growth curves shown in Fig. \ref{figApp2.1}. The median growth curve is given by $\rm a_0+ a_1 x + a_2 x^2 + a_3 x^3 + a_4 x^4 + a_5 x^5$, where $\rm x=R/R_e$.}
	\label{table_app_corr}
	\centering
	\begin{tabular}{ l c c c c c c }
		 & $\rm a_0$ & $\rm a_1$ & $\rm a_2$ & $\rm a_3$ & $\rm a_4$ & $\rm a_5$ \\
		\hline 
		$\rm \frac{SFR(<R)}{SFR(<2.5 R_e)}$  & 0.025 & 1.666 & 1.133 & -0.878 & 0.240 & -0.023  \\
	    $\rm \frac{ M_\star(<R)}{M_\star(<2.5 R_e)} $ & 0.018 & 0.586 & 0.455 & -0.544 & 0.191 & -0.023  \\
	\end{tabular}
\end{table}

For each MaNGA galaxy we calculate the integrated SFR and $\rm M_\star$ by summing the spaxel values in the associated maps within either $\rm 2.5 \ R_e$ or the largest radius at which the galaxy is observed by the MaNGA bundle, whichever is smaller. If the MaNGA bundle does not cover the galaxy to $\rm 2.5 \ R_e$ we use the fits to the growth curves above to compute a correction. Overall, 60\% of our sample does not require a correction, while the average correction for the rest of the sample is 0.04 dex for SFR and 0.05 dex for $\rm M_\star$. 

The stellar masses computed in this work within $\rm 2.5 \ R_e$ are not directly comparable to the stellar masses derived from integrated photometry, because a non-negligible fraction of a galaxy's stellar mass may lie outside $\rm 2.5 \ R_e$. For canonical exponential and de Vaucouleur profiles a $\rm 2.5 \ R_e$ aperture collects 92\% and 75\% of the light respectively. We do not attempt to perform further aperture correction to bring our masses in agreement with those measured from photometry, but we nonetheless compare the two to demonstrate the quality of the overall agreement. In this comparison all stellar masses have been converted to a Chabrier IMF and we consider the subsample of 579 MaNGA galaxies used to derive radial profiles in this work.

\begin{figure} 
	
	\includegraphics[width=0.42\textwidth, trim=10 0 0 20, clip]{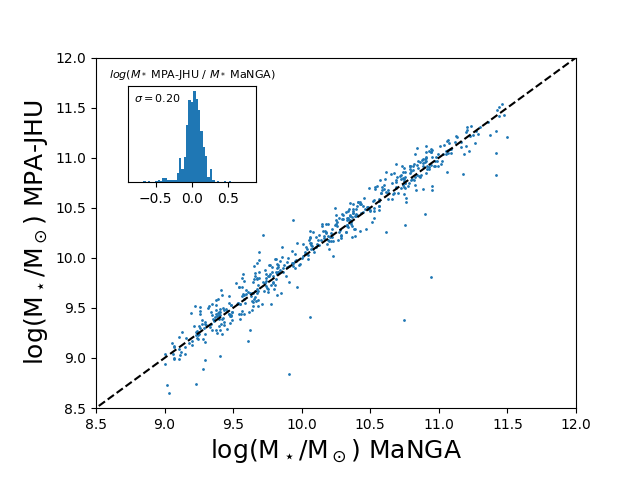}
	\includegraphics[width=0.42\textwidth, trim=10 0 0 20, clip]{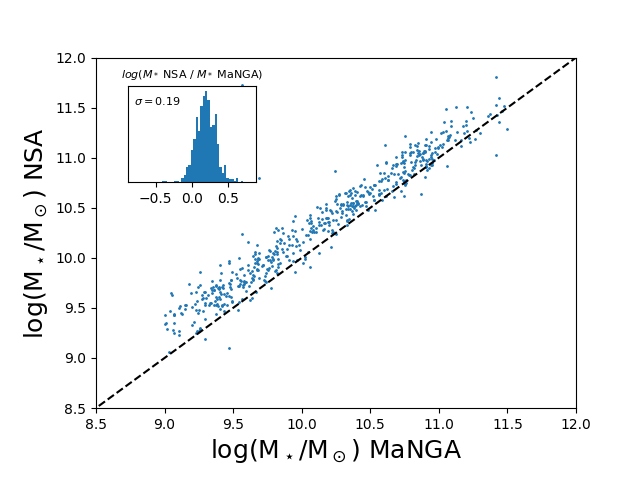}
	\caption{\textit{Top}: Comparison between the stellar masses from the MPA-JHU catalogue and the stellar masses derived in this work from the MaNGA data (integrated within $\rm 2.5 \ R_e$). \textit{Bottom}: Comparison between the stellar masses from the NSA catalogue (using S\'ersic photometry) and the stellar masses derived in this work from the MaNGA data (integrated within $\rm 2.5 \ R_e$).} 
	\label{figApp0}
\end{figure}

We first consider the stellar masses from MPA-JHU catalogue, based on SDSS DR8\footnote{http://www.sdss3.org/dr8/spectro/galspec.php}, which have been calculated based on $ugriz$ photometry following the methodology described in \cite{Kauffmann2003b}. The MPA-JHU stellar masses agree well with the MaNGA-derived one (offset of 0.03 dex) and show a relative scatter of 0.20 dex (Fig. \ref{figApp0}, top). There are some notable outliers from the 1-1 relation, with MPA-JHU masses up to 1 dex lower than MaNGA masses. If the MPA-JHU masses were correct, these galaxies would have unrealistically low i-band mass-to-light ratios. We therefore speculate that the MPA-JHU masses for these objects are incorrect. Their exclusion from the comparison lowers the scatter between MaNGA and MPA-JHU to 0.14 dex.

We also compare the MaNGA stellar masses with the stellar masses derived in the NSA catalogue using the k-correction procedure of \cite{Blanton2009}. The NSA stellar masses are on average 0.20 dex higher than the MaNGA-derived ones and the scatter between the two sets of mass measurements is 0.19 dex (Fig. \ref{figApp1}, top). We note that in the NSA catalogue the galaxies identified as problematic in MPA-JHU do not appear as outliers. 

Considering the wide variety of systematic effects affecting stellar masses measurements it is difficult to comment on the possible severity of aperture effects (due to the finite size of the MaNGA bundle) on the MaNGA-derived stellar masses based on this comparison alone. We note, however, that the definition of the GV is very robust with respect to the absolute scale of the stellar mass axis. The slope of the SFMS would, on the other hand, be affected. If we consider the MaNGA-derive SFR and the MPA-JHU masses the derived slope of the SFMS is 0.71 (and scatter of 0.41 dex). If the NSA masses derived from S\'ersic photometry are utilised instead, we obtain a slope of 0.84 for the SFMS (and 0.39 dex scatter).

\section{The properties of quiescent regions in central LIER galaxies}
\label{app3}

In this work we have argued that cLIER galaxies host quiescent central regions, and that LIER excitation is a useful classifier to select these centrally quiescent objects. In the scenario where LIER emission is due to hot evolved stars, the EW(H$\alpha$) in emission is expected to be low, as indeed observed in the centres of cLIER galaxies. A complementary measure of quiescence may be derived from the stellar population properties, by analysing, for example, an age-sensitive tracer like $\rm D_N(4000)$. In this appendix we explicitly show the equivalence of these different definitions of quiescence, in accordance with the detailed analysis already presented in \cite{Belfiore2016a} and \cite{Belfiore2017}.

\begin{figure} 
	
	\includegraphics[width=0.52\textwidth, trim=10 0 0 20, clip]{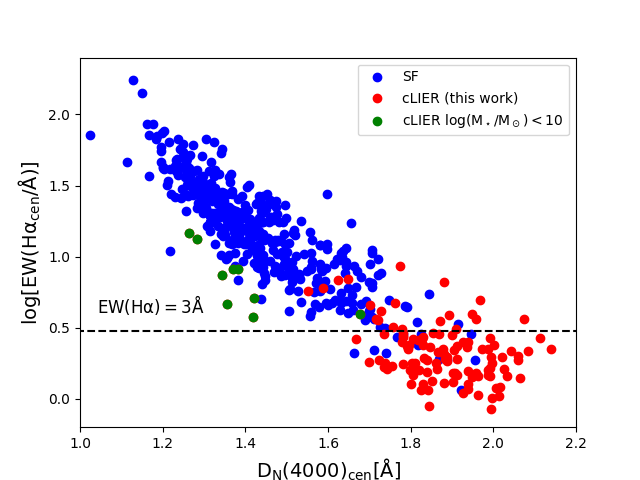}
	\caption{The position of star forming (blue) and cLIER galaxies used in this work ($\rm log(M_\star/M_\odot) > 10.0$, red points) in the EW(H$\alpha$) - $\rm D_N(4000)$ plane. EW(H$\alpha$) and $\rm D_N(4000)$ are computed from the MaNGA data using a central aperture 3$''$ in diameter. The dotted line represents the empirical division between star forming and passive galaxies (EW(H$\alpha$) $=$ 3\AA) suggested in previous SDSS work \protect\citep{CidFernandes2011}. Low-mass cLIER galaxies ($\rm log(M_\star/M_\odot) <10.0$) are also shown as green points. These galaxies are low-mass systems where the central aperture falls in an area dominated by diffuse ionised gas and have different properties from the high-mass cLIER galaxies, where the central regions are classified as quiescent using both EW(H$\alpha$) and $\rm D_N(4000)$.} 
	\label{figApp3}
\end{figure}

First, we focus on the properties of the central regions of galaxies. For each galaxy in our sample we derive the mean EW(H$\alpha$) and $\rm D_N(4000)$ in a central aperture 3$''$ in diameter. In Fig. \ref{figApp3} we show the position of star forming (blue) and cLIER with $\rm log(M_\star/M_\odot) >10.0$ (red) galaxies in the plane defined by the central EW(H$\alpha$) and $\rm D_N(4000)$. We see that cLIER galaxies lie at the high $\rm D_N(4000)$ and low EW(H$\alpha$) end of the distribution. An EW(H$\alpha$) of 3\AA\ is found to be a good empirical division between star forming and passive galaxies in SDSS \citep{CidFernandes2011}. In the current sample, 84\% of high-mass cLIER have EW(H$\alpha$) < 3 \AA\ and 96\% have  EW(H$\alpha$) < 6 \AA. We conclude that the central regions of high-mass cLIERs are classified as quiescent using both the EW(H$\alpha$) and $\rm D_N(4000)$.

The position of the 9 low-mass cLIERs $\rm log(M_\star/M_\odot) <10.0$, which have not been used in this work, is also shown in Fig. \ref{figApp3} (green dots). These galaxies lie in the region of the  EW(H$\alpha$) - $\rm D_N(4000)$ occupied by the star forming galaxies, although they occupy the lower EW(H$\alpha$) tail of the distribution. As already noted in the text, these galaxies are mostly irregular galaxies with clumpy line emission, and do not correspond to galaxies having old and quiescent central regions.

\bsp
\label{lastpage}

\end{document}